\documentclass[aps,prd,showpacs,preprintnumbers,twocolumn,groupedaddress,nofootinbib]{revtex4}
\usepackage{graphicx}
\usepackage{latexsym}
\usepackage{booktabs}
\usepackage{multirow}
\usepackage{amsmath, footnote, epsfig}
\pdfoutput=1

\def\beq{\begin{equation}}
\def\eeq{\end{equation}}
\def\bey{\begin{eqnarray}}
\def\eey{\end{eqnarray}}

\def\lsim{\mathrel{\raise.3ex\hbox{$<$\kern-.75em\lower1ex\hbox{$\sim$}}}}
\def\gsim{\mathrel{\raise.3ex\hbox{$>$\kern-.75em\lower1ex\hbox{$\sim$}}}}

\newcommand\T{\rule{0pt}{2.6ex}}       % Top strut
\newcommand\B{\rule[-1.2ex]{0pt}{0pt}} % Bottom strut

\begin{document}

\title{The Anisotropy of the Extragalactic Radio Background from Dark Matter Annihilation}
%\title{The Expected Anisotropy of Synchrotron Radiation from Extragalactic Dark Matter Annihilation}
\author{Ke Fang, Tim Linden}
\affiliation{Kavli Institute for Cosmological Physics, University of Chicago, Chicago, IL}
%\pacs{95.35.+d;95.30.Cq,95.55.Ka; FERMILAB-PUB-10-472-A}
\begin{abstract}
Observations of the extragalactic radio background have uncovered a significant isotropic emission across multiple frequencies spanning from 22~MHz to 10~GHz. The intensity of this non-thermal emission component significantly exceeds the expected contribution from known astrophysical sources. Interestingly, models have indicated that the annihilation of dark matter particles may reproduce both the flux and spectrum of the excess. However, the lack of a measurable anisotropy in the residual emission remains challenging for both dark matter and standard astrophysical interpretations of the ARCADE-2 data.  We calculate the expected synchrotron anisotropy from dark matter annihilation and show that these models can produce very small anisotropies, though this requires galaxy clusters to have large substructure contributions and strong magnetic fields. We show that this constraint can be significantly relaxed, however, in scenarios where electrons produced via dark matter annihilation can be efficiently reaccelerated by Alfv\'en waves in the intra-Cluster medium. Our analysis indicates that any source capable of explaining the intensity and isotropy of the extragalactic radio excess must have a spatial extension far larger than typical for baryons in galaxies, hinting at a novel physics interpretation.
\end{abstract}
\maketitle

\section{Introduction}
\label{sec:introduction}

%What is the ARCADE-2 Observation
Recently, the ARCADE-2 (Absolute Radiometer for Cosmology, Astrophysics and Diffuse Emission) collaboration reported an excess in the absolute temperature of the diffuse isotropic sky at frequencies between 0.022-90~GHz~\citep{arcade_interpretation}. This analysis utilized data taken with both the ARCADE-2 instrument between 3-90 GHz~\citep{arcade_measurement} as well as low-frequency observations including those reported by \citep{roger_22MHz_excess, guzman_44Mhz_excess, haslam_408Mhz_excess, reich_1.4GHz_excess}. After subtracting emission from the cosmic microwave background, the isotropic residual is found to have an absolute temperature falling as T($\nu$)~$\propto$~$\nu^{-2.6}$, with detectable emission at frequencies as high as 10~GHz. 

%Why galaxies don't produce this emission
This excess emission is difficult to explain with known astrophysical mechanisms. Models of the isotropic emission from the Milky Way predict an intensity that is a factor of a few too small to explain the observed signal. Specifically, these galactic components are constrained by null observations in CII H$\alpha$ and H2 studies~\citep{kogut_excess_not_galactic, arcade_measurement}. Extrapolations of the luminosity function of extragalactic source populations, such as radio galaxies, fail to reproduce the total intensity of the ARCADE-2 signal by a factor of 3-6~\citep{singal_not_extragalactic_baryonic_signals, gervasi_radio_luminosity_from_extrapolation_of_known_sources}. Additional astrophysical models were also investigated, including: (1) the diffuse backgrounds due to the propagation of free electrons, (2) additional low surface brightness objects, (3) radio supernovae, and (4) radio-quiet quasars. All of these models were found to be incompatible with the observed excess  \citep{singal_not_extragalactic_baryonic_signals}.

%Why dark matter can
Dark matter annihilation may provide an additional source of electrons and positrons capable of powering the ARCADE-2 emission through synchrotron emission~\citep{fornengo_arcade_excess_is_dm}. This explanation is motivated by the fact that dark matter annihilation is expected to occur in all high density regions, producing a relatively isotropic synchrotron signal over the extragalactic sky. Additionally, since dark matter annihilation can produce copious electrons without a corresponding thermal emission, dark matter annihilation naturally explains the high ratio of radio to IR intensity -- an observation that is difficult to fit with baryonic emission mechanisms. Subsequent investigations determined that the emission intensity and spectrum resulting from the annihilation of relatively low mass (m$_\chi$~$\lsim$~100~GeV) dark matter provided a reasonable fit to the ARCADE-2 excess. These classes of models are additionally motivated as an explanation to the $\gamma$-ray excess observed near the galactic center of the Milky Way~\citep{Goodenough:2009gk, hooper_goodenough, hooper_linden_gc, Abazajian:2012pn, Hooper:2013rwa, Gordon:2013vta, Macias:2013vya, Abazajian:2014fta, Daylan:2014rsa, Abazajian:2014hsa, Calore:2014xka}.

%Why dark matter can't (signal is to anisotropic)
However, a recent analysis by \citet{holder_anisotropy_of_arcade} employed observations by the Australia Telescope Compact Array (ATCA) and the Very Large Array (VLA) at  $4.86\,\mathrm{GHz}$, $8.4\,\mathrm{GHz}$ and  $8.7\,\mathrm{GHz}$ ~\citep{formalont_extragalactic_anisotropy_4.85GHz, partridge_extragalactic_anisotropy_8.4GHz, subrahmanyan_extragalactic_anisotropy_8.7GHz}, and showed that the anisotropy of the excess emission is significantly lower than expected from any source population that traces large scale structure. This observation challenges both dark matter and standard astrophysical explanations for the ARCADE-2 data. Notably, in order for a source population tracing large scale structure to have the anisotropy measured in VLA observations ($\triangle T / T_{\mathrm{CMB}}$~$<$~1.4~$\times$~10$^{-5}$ at 8.7 GHz), the emission must originate before z~$\approx$~5~\citep{Cline:2012hb}.

%Why this measurement is not necessarily correct and a better one is needed
It is not trivial, however, to directly translate the anisotropy of large scale structure to the expected anisotropy of synchrotron radiation produced via dark matter annihilation. For example, it was pointed out by \citet{holder_anisotropy_of_arcade} that sources with significant spatial extension ($\gsim$~1') may smear out the observed anisotropy in high angular resolution radio surveys. In addition, the luminosity spectrum of dark matter emission halos (dN/dL) may not correlate with the baryonic density, producing a very different anisotropy signature compared to galactic structure. 

Additionally, a recent analysis by \citet{Vernstrom:2014uda} examined the 1.75~GHz sky found that the sky temperatures could not be produced by any population of sources with individual fluxes greater than 1~$\mu$Jy and sizes smaller than 2$'$. Using these results, they argue that the ARCADE-2 emission is difficult to explain with any known source class, requiring an extremely large population of very faint radio emitters. Specifically, the models of  \citep{Vernstrom:2014uda} rule out emission from AGN, starburst galaxies, and radio galaxies, and appear inconsistent with models of dark matter annihilation.

%What we do (provide that better measurement)
In this \emph{paper}, we directly compute the anisotropy and source-flux distribution of the synchrotron signal from dark matter annihilation, under a wide variety of assumptions. The null observations of anisotropy from the extragalactic radio background strongly constrain many dark matter scenarios for the ARCADE-2 excess, forcing our models into a regime of parameter space where the magnetic field of galaxy clusters has a relatively large spatial extent, and the dark matter annihilation rate in clusters contains a significant contribution from dark matter substructure. However, we find regions within this parameter space where dark matter annihilation can produce the intensity and spectrum of the ARCADE-2 excess while remaining consistent with the radio isotropy and the paucity of bright point sources. In addition, we discuss an alternative model that could avoid large substructure contributions and strong magnetic fields by invoking the reacceleration of electrons via Alfv\'en waves powered by cluster mergers. Intriguingly, our results show that any emission mechanism aiming to explain the ARCADE-2 emission without overproducing the observed anisotropy must be dominated by arcminute-scale objects, rather than galactic emission. Noting that there is no baryonic mechanism that is known to produce smooth radio emission on cluster-sized scales with negligible galactic emission, we argue that the low-anisotropy of the ARCADE-2 excess instead hints at a dark matter explanation. 

%Outline of Paper, "In Section X"
The outline of the paper is as follows. In Section~\ref{sec:Cl_limits} we note that the constraints on radio anisotropy produced by \citet{holder_anisotropy_of_arcade} can fluctuate significantly based on uncertainties in the intensity of the ARCADE-2 emission at high frequencies. In Section~\ref{sec:models} we produce a complete model for the anisotropy of synchrotron radiation from dark matter annihilation. In Section~\ref{sec:results} we calculate the expected radio anisotropy of dark matter annihilation under a multitude of model assumptions. In Section~\ref{sec:Alfven} show that the reacceleration of electrons by Alfv\'en waves in cluster shocks can significantly enhance the dark matter synchrotron signal. Finally, in Section~\ref{sec:conclusions} we describe the implication of our results for interpretations of the ARCADE-2 excess.

\section{The Measured Anisotropy of the ARCADE-2 Excess}
\label{sec:Cl_limits}

At high frequencies ($\gsim$~1~GHz), the flux from the ARCADE-2 excess forms a subdominant contribution to the observed antenna temperature, making it difficult to directly measure the anisotropy of the excess. Fortunately, the total radio power is dominated by the CMB, which is highly isotropic compared to large scale structure. This makes it likely that the ARCADE-2 signal dominates the total temperature anisotropy. Using measurements from the Very Large Array (VLA) at frequencies of  $4.86\, \mathrm{GHz}$ and  $8.4\,\mathrm{GHz}$ \cite{formalont_extragalactic_anisotropy_4.85GHz, partridge_extragalactic_anisotropy_8.4GHz}, and the Australian Telescope Compact Array (ATCA) at $8.7\,\mathrm {GHz}$ \cite{subrahmanyan_extragalactic_anisotropy_8.7GHz}, \citet{holder_anisotropy_of_arcade} converted the limits from the measured temperature variance  $\triangle T / T_{\,\mathrm {CMB}}$ into an upper limit on the anisotropy of the ARCADE-2 emission using the relation:

\begin{equation}
\frac{\triangle T}{T_{\mathrm {excess}}} =  \frac{\triangle T}{T_{\mathrm{CMB}}} \,\frac{T_{\mathrm {CMB}}}{T_{\mathrm {excess}}}
\end{equation} 
where $T_{\mathrm {excess}}$ is the  temperature of the residual background from unresolved radio sources:
\begin{equation}
T_{\mathrm {excess}} = T_{\mathrm {arcade}} - T_{\mathrm {counts}}
\end{equation}

$T_{\mathrm{ counts}}$ is the expected temperature of the radio sky by extrapolation of known source counts \citep{gervasi_radio_luminosity_from_extrapolation_of_known_sources}:
\begin{equation}
T_{\mathrm {counts}} = 0.231\pm0.007\,\left(\frac{\nu}{\mathrm {GHz}}\right)^{-2.707\pm0.027}\,\mathrm {K}
\end{equation}
and $T_{\mathrm arcade}$ is the radio temperature reported by the ARCADE-2 collaboration \cite{arcade_interpretation}. Finally, the upper limits on $C_l$ is simply calculated by:

\begin{equation}\label{eqn:Cl_limits}
\frac{C_\ell\,\ell\,(\ell+1)}{2\pi}=\left(\frac{\triangle T}{T_{\mathrm {excess}}}\right)^2
\end{equation} 

which can be directly compared to theoretical models of the synchrotron anisotropy from dark matter annihilation. In \citet{holder_anisotropy_of_arcade},  $T_{\mathrm{arcade}}$ was calculated using an analytic spectrum fit to the ARCADE-2 data \citep{arcade_interpretation}, 
\begin{equation}
T_{\mathrm {fit}} = 1.483\pm0.100\,\left(\frac{\nu}{\,\mathrm{ GHz}}\right)^{-2.599\pm0.036}
\end{equation}

However, we note that this interpolation relies on the assumption that the emission producing the ARCADE-2 excess has a power-law spectrum, which may be violated in many dark matter models. Notably, there are significant uncertainties in the intensity of the ARCADE-2 excess above $\sim4 \,\mathrm{GHz}$, allowing for a temperature excess that falls below the value predicted by $T_{\mathrm {fit}}$ by more than an order of magnitude. These temperature uncertainties propagate into uncertainties in the synchrotron anisotropy ($C_\ell$) as outlined above, allowing for significantly higher anisotropies in the cases where T$_{\mathrm {excess}}$ has been overestimated. For the remainder of this work, we will directly compute the anisotropy constraints independently for each proposed model based on the excess temperature from dark matter synchrotron predicted in the VLA and ACTA bands by each individual model.

\section{Models for the Synchrotron Anisotropy from Dark Matter Annihilation}
\label{sec:models}

%The model of Fornengo et al. 
A detailed mathematical model for calculating the anisotropy of extragalactic dark matter annihilation has already been developed by \citep{ando_komatsu_anisotropy_2006} (see also \citep{ando_komatsu_anisotropy_2013}) and has been applied to models of the dark matter synchrotron anisotropy by \citep{fornengo_arcade_excess_is_dm2}. In this work, we will follow this mathematical model closely, with the exception of several alterations which will be noted in this section. At energy $E_s$,  the mean intensity and angular power spectrum of the synchrotron emission  from dark matter annihilation can be written as:

\begin{equation}\label{eqn:intensity}
I(E_{\mathrm{s}}) \equiv \frac{dN}{dE_sdtdAd\Omega}= \int d\chi  \delta^2(z)\,W[(1+z)E_s, \chi]
\end{equation}
\begin{equation}\label{eqn:anisotropy}
C_\ell (E_s)= \frac{1}{ I(E_s)^2}  \int \frac{d\chi}{\chi^2}W^2[(1+z)E_s, \chi]\,P_{\delta^2}(k,z)
\end{equation}
where k=$\ell$/$\chi$, $\chi$ is the comoving distance at redshift $z$ ($d\chi/dz = c/H(z)$ with $H$ the Hubble parameter and $c$ the speed of light). 
The clumping factor $\langle\delta^2(z)\rangle$ is the variance of the dark matter overdensity, $\delta\equiv (\rho-\langle \rho\rangle)/\rho$, and is  given by:

\begin{equation}
\langle \delta^2(z) \rangle =\int_{M_{min}}^\infty dM \frac{dn(M, z)}{dM} \int dV u(r, M,z)
\end{equation}
where $u$ is the square of the density profile of a dark matter halo of mass $M$ at redshift $z$:
\begin{equation}
u \equiv \rho^2(r, M, z) / \left(\Omega_{\rm DM}\rho_c\right)^2
\end{equation}

The window function W encodes the emission intensity stemming from dark matter annihilation:

\begin{equation}
W[E_s, z] = \frac{\langle\sigma v\rangle}{8\pi}\left(\frac{\Omega_{\mathrm {DM}}\rho_c}{M_{\mathrm {DM}}}\right)^2 (1+z)^3~\frac{dN_s}{dE_s}e^{-\tau(z)} 
\end{equation}

%\setlength{\thinmuskip}{0mu}
%\begin{equation}
%C_\ell = \left(\frac{\sigma v}{8\pi}\right)^2\left(\frac{\rho_\chi}{M_\chi}\right)^4 \int \frac{c(1+z)^6~dz}{H(z)r^2(z)}~\left(\frac{dN_s}{dE}\right)^2P(k, z)e^{-2\tau(z)} 
%\end{equation}
%\setlength{\thinmuskip}{2mu}

where $\langle\sigma v\rangle$ is the thermally averaged cross-section times velocity for dark matter annihilation,  $\rho_{c}$ is the critical density,  $\Omega_{\mathrm {DM}}$ is the fractional density of dark matter in the universe, M$_{\mathrm {DM}}$ is the mass of the dark matter particle, and $dN_s/dE_s$ is the synchrotron emission intensity  for a given dark matter density profile which we will discuss in Sec.~\ref{subsec:syn}. 

Finally,  $\tau(z)$ is the optical depth of radio emission as a function of redshift. We note that for radio emission above 10~MHz, the universe is roughly radio transparent  \citep{1992ApJ...396..487S}, and we will neglect this parameter in what follows (set $\tau(z) = 0$). We begin our integration at z=0.001 in order to remove nearby sources~\citep{ando_komatsu_anisotropy_2006}. The upper limit should be taken to correlate with the maximum redshift for the formation of large scale structure. In our calculations we take an upper limit z$_{\mathrm {max}}$~=~5. Practically, however, the signal is dominated by emission within z~$<$~1, and thus the exact choice of z$_{\mathrm {max}}$ is immaterial. The cosmological parameters used in this work are taken from \citet{2011ApJS..192...18K}, which were calibrated by combining the WMAP data with the  distance measurements from the baryon acoustic oscillations (BAO) in the distribution of galaxies and the Hubble constant ($H_0$) measurement. 

\subsection{The Synchrotron Anisotropy Calculation}\label{subsec:Cl_calculation}

In Equation~\ref{eqn:anisotropy}, P(k,z) is the power spectrum of the dark matter overdensity squared, encoding all morphological information about the size and luminosity distribution of dark matter halos. The term P(k,z) can be broken down into a one-halo and two halo term  \citep{ando_komatsu_anisotropy_2006}, 
\begin{equation}
P(k,z) = P_{1h}(k,z) + P_{2h}(k,z)
\end{equation}
where the one-halo term denotes correlations between particles within the same halo,
\begin{equation}
P_{1h} (k,z) = \int dM \frac{dn}{dM}|\tilde u(k,M)|^2
\end{equation}
and the two-halo term denotes correlations between particles in two distinct halos:
\begin{equation}
P_{2h} (k,z) =  \left(\int \frac{dn}{dM} b(M, z)|\tilde u(k,M)|\right)^2 P^{\mathrm {lin}}(k,z)
\end{equation}
Here $\tilde u(k, M)$ is the Fourier transform of $u(r, M)$,  dn/dM is the halo mass function for primary halos, b(M,z) is the linear biasing term which describes the source clustering, and P$^{\mathrm {lin}}$ is the linear matter power spectrum given by the primordial linear power spectrum and multiplied by the transfer function. In our calculation we follow  \citet{2002PhR...372....1C} for $b(M,z)$ and \citet{1999ApJ...511....5E} for P$^{\mathrm {lin}}$.

The mass function, $dn/dM$ is given by:
\begin{equation}
\label{eq:dndM}
\frac{dn}{dM} = f(\sigma) \frac{\rho_m}{M}\frac{d \ln~\sigma^{-1}}{dM}
\end{equation} 
where $\rho_m$ is the mean density of the universe at the epoch of analysis, $\rho_m(z) =\rho_m(0)\,(1+z)^3$, and $\sigma(M, z)$   is the rms variance of the linear density field smoothed on a top-hat window function $R=(3M/4\pi\rho_m)^{1/3}$,
\setlength{\thinmuskip}{0mu}
\begin{equation}
\label{eq:sigma}
\sigma^2 = \int  dk\, P^{\mathrm {lin}}(k) \tilde W(kR)\,k^2
\end{equation}
where $\tilde W(kR)$  is the Fourier transform of the real-space top-hat window function of radius R \citep{2008ApJ...688..709T}. In our work we compute $\sigma(M, z)$ with CosmoloPy \footnote{http://roban.github.com/CosmoloPy/}. For $f(\sigma)$ we use the mass function multiplicity described by  \citet{sheth_tormen_mass_function}, which is a modification of the standard Press-Schechter mass-function~\citep{press_schechter_mass_function}.

Electrons and positrons produced via dark matter annihilation then diffuse and produce synchrotron radiation in the halo magnetic field, or up-scatter CMB photons to $\gamma$-rays by an inverse Compton process. Working under the assumption that these electrons cool faster than they appreciably diffuse (and thus the synchrotron signal from a given location is dependent only on the local annihilation rate)\footnote{Since e$^+$e$^-$ tend to diffuse from high density regimes to low density regimes, we note that this simplifying assumption overestimates the total anisotropy, and is thus highly conservative.}, the synchrotron intensity from dark matter annihilation can then be written as:

\begin{equation}
\rho_{\mathrm {sync}}^2(r,M) =\rho_{\mathrm {DM}}^2(r,M)S_{\mathrm{ eff}}(r,M)
\end{equation} 
where   
\begin{equation}\label{eqn:Seff}
S_{\mathrm {eff}} = \left(\frac{\rho_{\mathrm {mag}}}{\rho_{\mathrm{ mag}} + \rho_{\mathrm {CMB}}}\right)~\left(1-f_{\mathrm {esc}}(M,r)\right)~F_{s}
\end{equation}
where the first term is the ratio of the local magnetic field energy density to the CMB energy density, and specifies the fraction of the total lepton energy which is converted to synchrotron radiation as opposed to the inverse Compton scattering of CMB photons. Notably, the energy density of the CMB is equivalent to the energy density from a magnetic field of strength  B$\rm_{CMB}$ = $3.24\,(1+z)^2\,\mu$G. For magnetic field values larger than this, the majority of the electron energy will be converted to synchrotron radiation (with negligible improvements in total power for larger magnetic field values), while for magnetic fields below this limit the total synchrotron power will fall approximately as B$^{2}$.

The second term specifies the fraction of the lepton energy which is converted to synchrotron radiation before the relativistic leptons leave the cluster, and the last term $F_s$ gives the fraction of synchrotron emission which contributes at a specific frequency. This value has a radial dependence via the radial dependence of the halo magnetic field. We will show in Sec.~\ref{subsec:syn} that $f_{\mathrm {esc}}\sim 0$ and $F_s\approx 1$. 
 Note that in all cases 0~$\le$~S$_{\mathrm {eff}}$~$\le$~1. In the next few sections we will discuss our models for S$_{\mathrm {eff}}$ in great detail. 

\subsection{Dark Matter Density Profiles}
The local dark matter density profile, $\rho_{\mathrm {DM}}(r,M)$ is calculated using a generalized NFW profile~\citep{nfw_profile} given by:
\begin{equation}
\rho_{\mathrm {h}}(r,M) = \rho_0(M) \left(\frac{r}{r_{\mathrm{s}}}\right)^{-\gamma}\left(1+\frac{r}{r_{\mathrm{s}}}\right)^{-3+\gamma}
\end{equation}
where $\gamma$ is the inner slope of the dark matter profile, which is set to 1 in the default NFW analysis.  Utilizing recent evidence that the dark matter in galaxy sized objects may be adiabatically contracted near the galactic center, we additionally analyze models where $\gamma$~=~1.3 in objects 10$^{10}$-10$^{12}$M$_\odot$~\citep{gnedin_adiabatic_contraction}. However, we find that this has only a marginal effect on our results. The parameter r$_c$ is the core radius of the dark matter profile, and is calculated from the virial radius and the concentration parameter as r$_s$~=~r$_{\mathrm {vir}}/c_{\mathrm {vir}}$, and we use a redshift dependent model for the concentration parameter $c_{\mathrm {vir}}$ given by \citet{klypin_concentration_parameter}. The virial radius,  $r_{\mathrm {vir}}$, is determined through the relation: $M = 4\pi r_{\mathrm {vir}}^3\triangle_{\mathrm {vir}}(z)\,\rho_m(0)/3$, with $\triangle_{\mathrm {vir}} (z) \approx (18\pi^2 + 82 x - 39x^2)/(x+1)$ and $x =\Omega_M(z) - 1$ \citep{2003ApJ...584..702H}.
The value of $\rho_0(M)$ is then calculated to set the total mass M within the virial radius. For $\gamma=1$ this can be solved analytically~\citep{ando_komatsu_anisotropy_2013}:
\begin{equation}
\rho_0 = \frac{M}{4\pi r_s^3}\left[\ln (1+c_{\mathrm {vir}}) - \frac{c_{\mathrm {vir}}}{1+c_{\mathrm {vir}}}\right]^{-1}
\end{equation}

In addition to the smooth dark matter component given by the NFW profile, the annihilation rate of dark matter can be boosted by substructure in the dark matter distribution both inside and around the host halo. In our calculations, we adopt the probability distribution function of the dark matter profile in \citet{2010PhRvD..81d3532K}, which is calibrated to the results of the Via Lactea simulation \citep{2008Natur.454..735D}. Thus the mean density squared at given radius is calculated by:
\begin{eqnarray}
\frac{\langle \rho(r)^2\rangle}{\langle\rho(r)\rangle^2} &=& f_{\mathrm{s}}\,e^{\triangle^2} \\ \nonumber
&& + (1-f_{\mathrm{s}})\frac{1+\xi}{1-\xi}\left[\left(\frac{\rho_{\mathrm {max}}}{{\rho_{\mathrm{h}}}}\right)^{1-\xi}-1\right]
\end{eqnarray}
taking $\triangle\approx0.2$, $\xi\approx0$ and $\rho_{\rm max} = 80\,\rm GeV\,cm^{-3}$ for a galaxy-sized halo \citep{2010PhRvD..81d3532K}. The parameter $f_s$ is the  fraction of the dark matter mass in the smooth halo within the volume of the halo. Then $1-f_s$ is the fraction of the high-density clumped component, and is approximated by  \citep{2010PhRvD..81d3532K}
\begin{equation}
1-f_{\mathrm{s}}(r) = 7\times10^{-3}\left(\frac{\rho_{\mathrm{h}}(r)}{\rho_{\mathrm {h}}(r=100\,\mathrm{kpc})}\right)^{-0.26}
\end{equation}
This gives a total fraction of  $1\%$ for the high-density clumped component  in the volume of the halo, and an integrated boost factor of $20.5$ within the virial radius of a Milky Way-sized halo. We then scale the boost factor by $M_{\rm halo}^{0.39}$ for larger and smaller halos \citep{hooper_arcade_excess}. We note that the contribution from dark matter substructure can extend well beyond the virial radius of the host halo \citep{2008MNRAS.391.1685S, 2008ApJ...679.1260M, 2014ApJ...789....1D}. 
%For example, the study of \citet{} on the outer density profile of $\Lambda$CDM halos indicated that dark matter contribution can extend up to $9\,r_{\rm vir}$. 
In order to account for the contribution from substructure outside of the virial radius, we  extrapolate the density distribution function of \citet{2010PhRvD..81d3532K}  out to a radius $r_{\mathrm {sub}}= 1~-~10~\,r_{\mathrm {vir}}$. For a Milky Way sized halo, $r_{\mathrm {sub}} = 4~\,r_{\mathrm {vir}}$ would result an integrated boost  factor of $74.7$.
%, and $r_{\mathrm {sub}} = 10\,r_{\mathrm {vir}}$ would push the boost factor further to $163.1$. 
This boost factor is consistent with the results of the Aquarius simulation  \citep{2008MNRAS.391.1685S, 2012MNRAS.419.1721G, 2012MNRAS.425.2169G}, but is disfavored by  \citet{2014MNRAS.442.2271S}.

%\begin{equation}
%\label{eq:synchrotron_emission_equation}
%\frac{d\tilde N_{s}}{dE} = 2 \int_{m_e}^{M_\chi}\frac{P_s}{E}\tilde n_e
%\end{equation}

%to be the steady-state synchrotron emission intensity for a given dark matter injection rate where P$_s$ is the synchrotron emission power given by the formula

%\begin{equation}
%P_s(r,E,\nu) = \frac{\sqrt{3}e^3}{m_ec^2}B(M,r)F(\nu / \nu_c)
%\end{equation}

%where $\nu_c$ is the critical frequency for synchrotron emission, B(M, r) is the magnetic field, which we modify from \citet{hooper_arcade_excess} and \citet{dolag_magnetic_field_in_clusters} as:

\subsection{Magnetic Fields}\label{subsec:B}

In order to generate the observed synchrotron intensity, we must convolve our dark matter annihilation model with a magnetic field model for dark matter structures of different masses. In order to calculate the contribution from cluster-sized halos, which we define here to consist of halos greater than 2~$\times$~10$^{12}$~M$_\odot$, we adopt the same formalism as in \cite{hooper_arcade_excess} which gives the magnetic field as a function of the radial distance from the center of the cluster to be:

\label{subsec:bfield}
\begin{equation}\label{eqn:B_cluster}
B(M,r) = B_0 \left(\frac{M}{M_{0}}\right)^\alpha\left[1 + \left(\frac{r}{r_{\mathrm{c}}}\right)^2\right]^{-3\beta \eta/2}
\end{equation}

where B$_0$~=~40$\mu$G is our default magnetic field at a mass M$_0$ = 10$^{14}$M$_\odot$, $\alpha$ controls the dependence of the maximum magnetic field strength on the mass of the cluster in consideration, but is set to 0.0 in our default model. We take $3\beta$ to be the slope of the gas profile at r~$\gg$~$r_c$, with $\beta$~=~0.6, and $\eta$ to be the falloff of the magnetic field from the center of the distribution, which we take to be $\eta$~=~0.5. We define the core radius of the magnetic field model to be equivalent to r$_c$~=~0.05~R$_{vir}$ where R$_{vir}$ is the virial radius of the cluster. For galaxy-sized halos with magnetic fields dominated by their dense baryonic cores, we take a two part magnetic field model which adopts the larger of the two following magnetic fields:

\setlength{\thinmuskip}{2mu}
\begin{equation}\label{eqn:B_galaxy}
B(r) =\max\left( B_1 \,e^{-r/R_1},  B_2\,e^{-r/R_2}\right)
\end{equation}

where B$_1$ = 7.6 $\mu$G and R$_1$ = 0.025~R$_{\mathrm{vir}}$ and B$_2$~=~35~$\mu$G with R$_2$ = 0.008R$_{\mathrm{vir}}$. The latter magnetic field is added to account for the large magnetic fields observed in the center of the Milky Way galaxy~\citep{crocker_minimum_50muG}. We note that these radial values are given by the best fit radial profiles to those models adopted in ~\citep{hooper_arcade_excess}. 

\begin{figure}
\centering
\epsfig{file=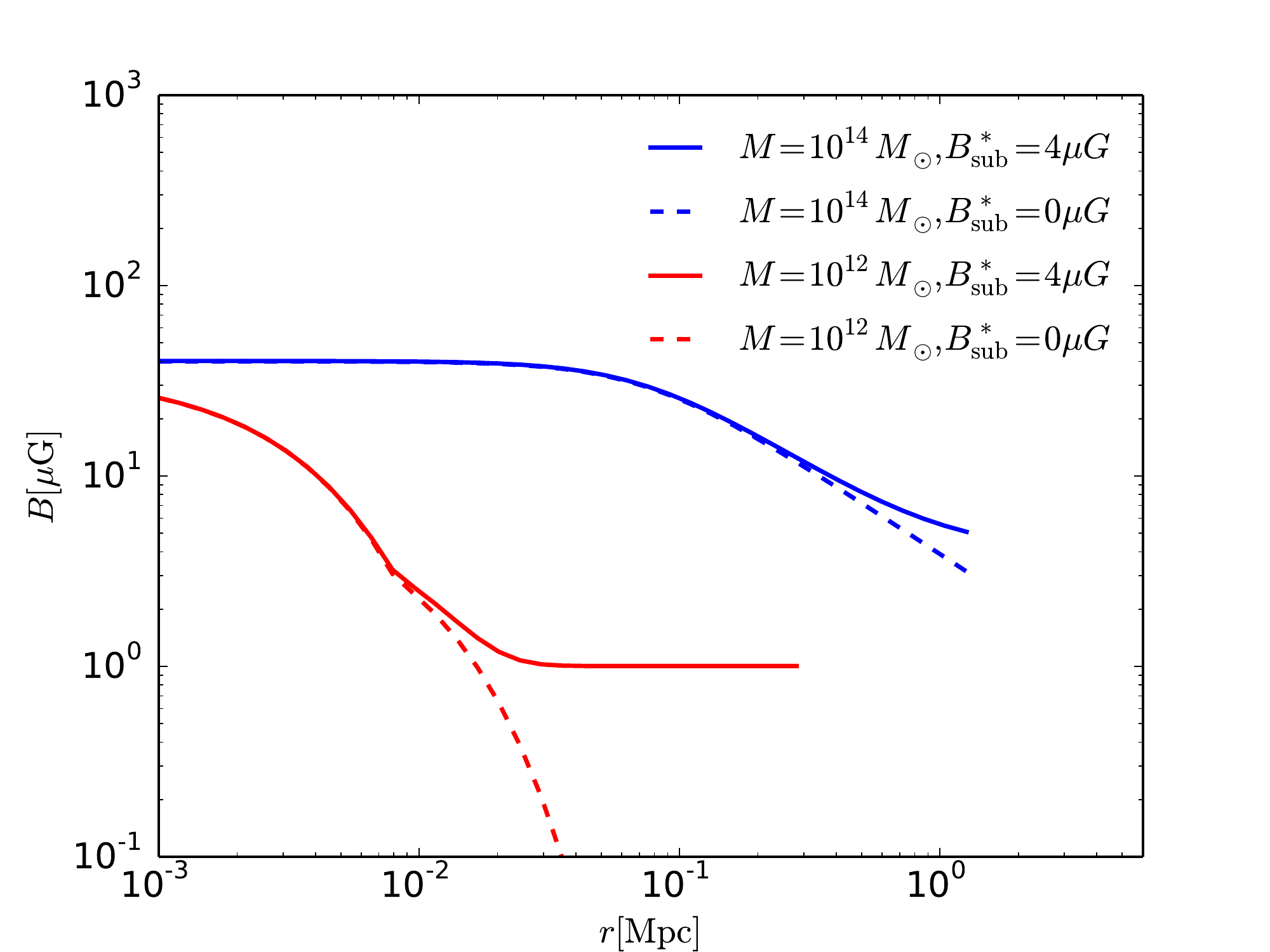,width=0.45\textwidth}  
\caption{\label{fig:B_profile} The radial distribution of the magnetic field in a $10^{14}\,M_\odot$ halo (blue) and a $10^{12}\,M_\odot$ (red) halo up to the virial radius. The dashed lines correspond to the  radial profiles of the best fit magnetic field models  adopted in  \citet{hooper_arcade_excess} (as demonstrated by Eqn.~\ref{eqn:B_cluster} for a cluster-sized halo and Eqn.~\ref{eqn:B_galaxy} for a galaxy-sized halo). The solid lines represent the magnetic field models used in  this work, where in addition to the best fit models, a lower-limit, $B^*_{\mathrm{sub}}=4\mu$G for a $10^{14}\,M_\odot$ halo (corresponding to $1\mu$G for a $10^{12}\,M_\odot$ halo) is set  to the field strength at large radii.}
\end{figure}

% B in substructure outside viriral radius
An important uncertainty concerns the extension of the magnetic field out to scales near the virial radius in cluster-sized objects. Recent studies have inferred magnetic field intensities between 0.1 and a few $\mu$G at the virial radius \citep{2002ARA&A..40..319C, 2009A&A...503..707B, 2012SSRv..166..187B}. Simulations of the cluster magnetic turbulence have suggested smaller values on the order of $0.1\,\mu$G, while observations of radio relics support magnetic field strengths that can be even larger than 3~$\mu$G  \citep{2006AJ....131.2900C, 2010Sci...330..347V}. Equipartition models suggest a still larger magnetic field, with an intensity of approximately 10 $\mu$G around the virial radius \citep{2005AN....326..414B}. In this work, we supplement the magnetic field models discussed above by setting a lower-limit, $B_{\rm sub}$,  on the magnetic field intensity that extends to the end-point of our simulation (at  $ r_{\rm sub}$). In particular, we take into account the mass dependence of this minimum magnetic field as:
\begin{equation}
B_{\rm sub} = B_{\rm sub}^*\, \left(\frac{M}{10^{14}\,M_\odot}\right)^\alpha
\end{equation}
where $B_{\rm sub}^*$ is minimum for a $10^{14}\,M_\odot$ halo, and in our calculation we take $\alpha = 0.3$.  
In different models we allow the intensity of this magnetic field component to fluctuate. In Fig.~\ref{fig:B_profile} we compare the radial profiles of the magnetic field models in a cluster-sized ($10^{14}\,M_\odot$)  halo and a galaxy-sized ($10^{12}\,M_\odot$) halo for $B_{\mathrm{ sub}}^* = 4\mu$G and $B_{\mathrm {sub}}^*=0$. The choice of $B_{\mathrm {sub}}^*$~$>$~0, produces strong magnetic fields at large radii, which may be at odds with cluster observations. However, these magnetic fields are included in order to probe the parameter space of dark matter fits to the ARCADE-2 excess. 

\subsection{Synchrotron Spectral Models}\label{subsec:syn}

The steady state distribution of the e$^+$e$^-$ injected via dark matter annihilation can be solved using the continuity equation \citep{hooper_arcade_excess}:
\begin{eqnarray}\label{eqn:continuity}
0 &=& \overrightarrow{\nabla}\cdot \left[D(\vec{x}, E_e)\overrightarrow{\nabla}\frac{dN_e}{dE_e}(\vec{x}, E_e)\right] \\ \nonumber
&+& \frac{\partial}{\partial E_e}\left[b(\vec{x}, E_e)\frac{dN_e}{dE_e}\right] + \left(\frac{dN_e}{dE_edt}\right)_{\rm inj}
\end{eqnarray}

The first term describes particle diffusion inside the cluster magnetic field. The diffusion coefficient for Kolmogorov diffusion in the Milky Way has been calculated to be approximately \citep{2012JCAP...01..010B}:
\begin{equation}
D(E)=2\times 10^{28}\,\left(\frac{E_e}{10\,\rm GeV}\right)^{1/3}\,\rm cm^2\,s^{-1}
\end{equation}

For this diffusion constant, the typical energy loss time scale from electron escape would be:
\begin{equation}
\tau_{\rm esc} \approx \frac{L^2}{D} = 63 \,\rm Myr 
\end{equation}
for a Milky Way sized halo with a typical diffusion scale of $L \sim 2\,\rm kpc$. In comparison,  the energy loss timescale for relativistic leptons from a combination of inverse Compton scattering and synchrotron emission occur on a characteristic timescale of:

\begin{equation}
\setlength{\thinmuskip}{1mu}
\tau_{\rm loss} \approx 21  \left(\frac{E_e}{10 \rm GeV}\right)^{-1}\left(\frac{B}{6 \mu \mathrm{G}}\right)^{-2}\,\rm Myr
\end{equation}

Thus, for galaxy sized halos, diffusion is not entirely negligible, and may decrease the total synchrotron power. However, in cluster sized halos, the diffusion scale, $L$, increases significantly, while the interstellar-radiation field and magnetic field energy densities are likely to increase,  making $\tau_{\rm loss} / \tau_{\rm esc} \ll 1$ and thus $f_{\rm esc}$~$\approx$~0. T. As most of the synchrotron signal in our model is contributed by cluster-sized halos, we neglect the effect of diffusion and simplify Eqn.~\ref{eqn:continuity} by setting the first term to zero.

%during this energy loss timescale, a particle is able to diffuse a distance
%\begin{equation}
%D = \sqrt{c \tau_{loss}}\ell
%\end{equation}
%where $\ell$ is the stepsize for diffusive motion which can be computed from the diffusion constant as:
%\begin{equation}
%\ell = \frac{6D_0}{c}
%\end{equation}

%here, we take D$_0$~=~5~$\times$~10$^{28}$~cm$^2$s$^{-1}$, and make a simplistic assumption that the diffusion constant is energy independent (which is approximately true for the very hard lepton energy spectra we are considering). In this case, we obtain $\ell$~=~1~$\times$~10$^{19}$~cm~=~3.24~pc, and thus we obtain a diffusion distance D~=~6.42~kpc during which 10~GeV lepton loses it's energy to the CMB in a magnetic field free environment. For the case where the particle is additionally in a 10$\mu$G magnetic field, we obtain D~=~2.4~kpc.

In this limit, a steady-state e$^+$e$^-$ spectrum can be calculated as:
\begin{equation}\label{eqn:steady_e}
\frac{dN_e}{dE_edV} = \frac{1}{b(E_e)}\,\int_{E_e}^{m_{\rm DM}}dE_e'\,\left(\frac{dN_e}{dE_e'dVdt}\right)_{\rm inj}
\end{equation}
where b is the energy dissipation rate through inverse Compton and synchrotron processes,
\begin{eqnarray}
b(E_e) = \frac{4}{3}\sigma_T\,c\,\left(\frac{E_e}{m_e}\right)^2\,(\rho_{\rm mag} + \rho_{\rm CMB})
\end{eqnarray}
and where $\sigma_T$ is the Thomson cross section.
Then the spectrum of synchrotron emission is given by
\begin{equation}
\frac{dN_s}{dE_sdtdV} = \frac{\sqrt{3}e^3B}{hm_e c^2\, E_s}\,\int_{m_e}^{m_{\rm DM}}dE_e\,F(\frac{E_e}{h\nu_c})\,\frac{dN_e}{dE_edV}
\end{equation}
%Notice that setting $1/dVdt$ in eqn~\ref{eqn:steady_e} to $\langle\sigma v\rangle\,(\rho_c\Omega_{\rm DM})^2$ we obtain $dN_s/dE_s$ for one annihilation.

where $F(x)=x\int_x^\infty K_{5/3}(x')dx'$ is the Synchrotron function and $K_{5/3}$ is the modified Bessel function. An accurate fitting formula of $F(x)$ \citep{2013RAA....13..680F} is used in our work, to increase processing speed. 

%We have, so far, ignored any discussion of the synchrotron spectrum, which is incorporated into the model of F12 as a term $\frac{dN_s}{dE}$, which plots the synchrotron spectrum from a dark matter annihilation in a given magnetic field. We note that the spectral fit of dark matter to the ARCADE-2 excess has already been established by \citet{fornengo_arcade_excess_is_dm} and \citet{hooper_arcade_excess}. Thus in this paper, we will not consider the spectral fit more closely, but will instead concern ourselves on with the fraction of 

%In this paper, we assume that the contribution to synchrotron at the observed frequency is independent of the magnitude of the magnetic field and the annihilation spectrum of the dark matter particle. We note the critical frequency for synchrotron emission is given by:

The critical frequency of synchrotron emission is determined by 
\begin{eqnarray}
\nu_c &=& \frac{3}{2}\left(\frac{E_e}{m_e}\right)^2\frac{eB}{2\pi m_e c } \\ \nonumber
&=&5.7\, (\frac{B}{5 \mu \rm G})(\frac{E_e}{10 \rm GeV})^2 ~ \rm GHz
\end{eqnarray}
The peak of the synchrotron emission spectrum occurs at $\nu$/$\nu_c$~$\sim$~0.29, which is approximately 2~GHz for an $10\,\rm GeV$ electron, near the frequency of present radio observations.  In this case, the synchrotron spectrum is relatively flat for radiation produced in regions with different magnetic field strengths. We find that the majority of our emission stems from regions where the magnetic field varies only between 5-10$\mu$G. This changes the frequency of the peak synchrotron emission by only a factor of two, leading to a smaller than 10\% change in the total synchrotron intensity at energies near the peak. Thus, in these models we assume that the magnetic field strength used to calculate dN$_s$/dE$_s$ is fixed to be $5\,\mu$G. In this case, the contribution of the synchrotron intensity to Eqn.~\ref{eqn:anisotropy} falls out entirely, as it is cancelled by the mean synchrotron intensity $\langle I_\gamma \rangle$. We will not consider this spectral term any further and in everything that follows we assume $F_s=1$ in Eqn.~\ref{eqn:Seff}.

\section{Results}
\label{sec:results}

\begin{table}[t]
\caption{Summary of Input Parameters} \label{table:input}
\centering
\begin{tabular}{ccccccccc}
\hline\hline 
Case &$m_{\rm DM}$ \T &~ annihilation~& $\langle\sigma v\rangle$ & $r_{\rm sub}$ & $B^*_{\rm sub}$  & $\chi^2\footnote{There are 10 degrees of freedom. We note that our calculation of $\chi^2$ removes the datapoint at 7.97~GHz, which has an extremely small statistical error, but appears to have systematic problems. This approach was advocated by the ARCADE-2 team~\citep{arcade_measurement}.}$\\ &   (GeV) &~channel~& ($\rm cm^3s^{-1}$)& ~($r_{\rm vir}$) & ~($\mu$G)~ & &  \B  \\
\hline
I & 50 & $b\bar{b}$ & $3\times10^{-26}$ & 8 & 8 &72.64\B\\
II \T & 8 &  leptons & $8.4\times10^{-27}$ & 4 & 4 & 44.58 \\
III & 23 & charge coupled  & $7.2\times10^{-27}$ & 8 & 8 & 56.65\\
\hline
\end{tabular}
\end{table}

A thorough examination of the spectral models which fit the ARCADE-2 excess has already been presented in \citet{hooper_arcade_excess}. However, different spectral models also produce different anisotropies through their effect on the total dark matter flux above 4~GHz in models that are similarly normalized to fit the ARCADE-2 excess at lower frequencies.  In this work we focus on relatively low-mass dark matter models for two particular reasons. First, low-energy electrons efficiently produce synchrotron emission at low-frequencies while producing a smaller fraction of high frequency synchrotron radiation. This has the effect of decreasing the necessary electron production rate and decreasing the anisotropy in high frequency observations, as shown in Section~\ref{sec:Cl_limits}. Second, these models are well motivated by their ability to explain the $\gamma$-ray signal observed in the Galactic Center by the Fermi-LAT telescope~\citep{Goodenough:2009gk, hooper_goodenough, hooper_linden_gc, Abazajian:2012pn, Hooper:2013rwa, Gordon:2013vta, Macias:2013vya, Abazajian:2014fta, Daylan:2014rsa, Abazajian:2014hsa, Calore:2014xka}. 

\subsection{Dark Matter Models}

In Table~\ref{table:input} we list the input parameters of three different base models. For each case, we have allowed the annihilation cross-section ($<$$\sigma v$$>$), the maximum radius of our simulation (r$_{\rm sub}$) and the maximum magnetic field strength which extends to the limit of the simulation (B$_{\rm sub}$) to float in order to achieve the best fit to the temperature spectrum without overproducing the observed anisotropy. We will now discuss each case independently, in light of both its feasibility as a model for the ARCADE-2 excess, and also in light of constraints from other indirect detection studies. 

\begin{figure}
\centering
\epsfig{file=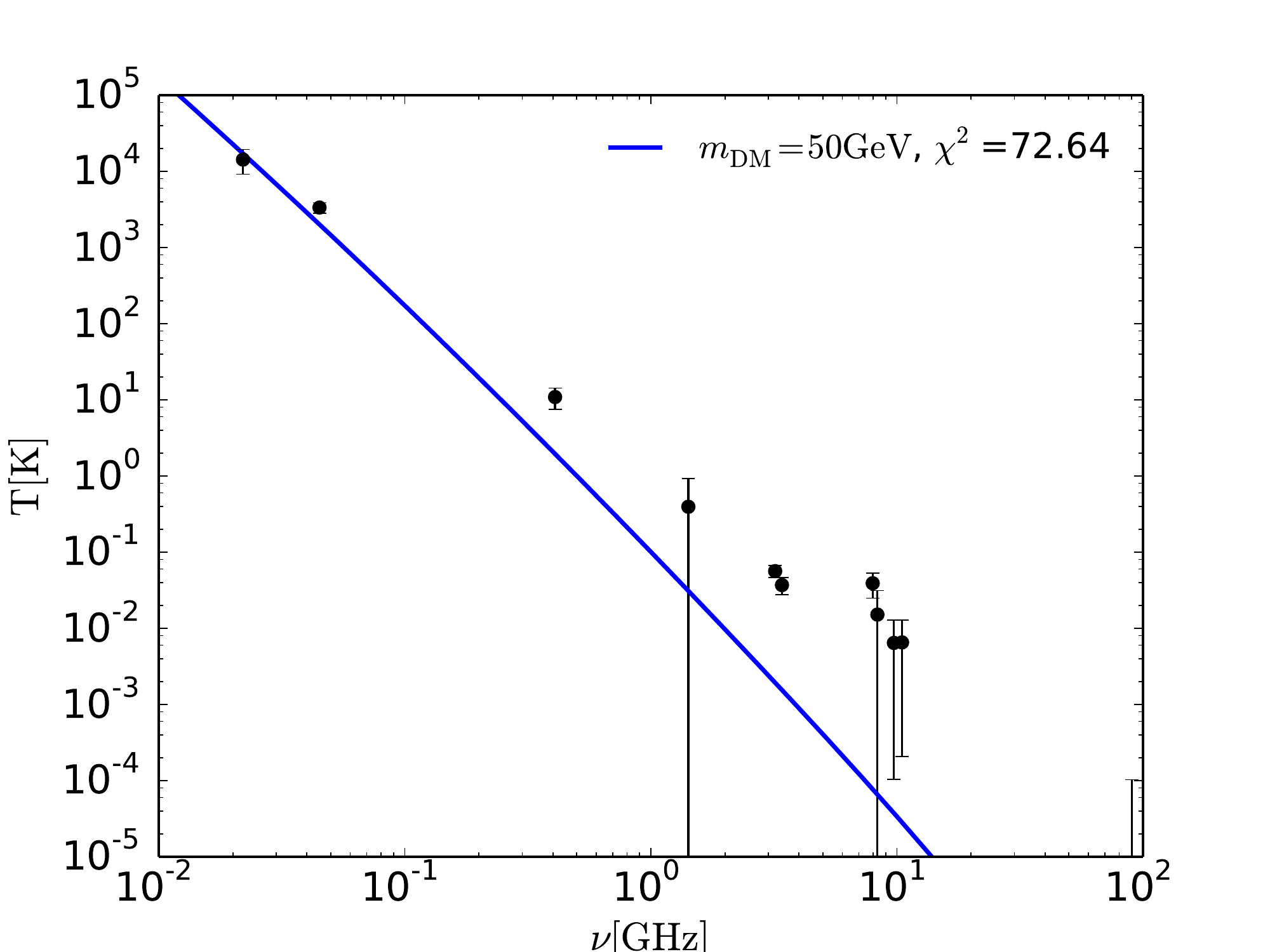,width=0.45\textwidth}   
\epsfig{file=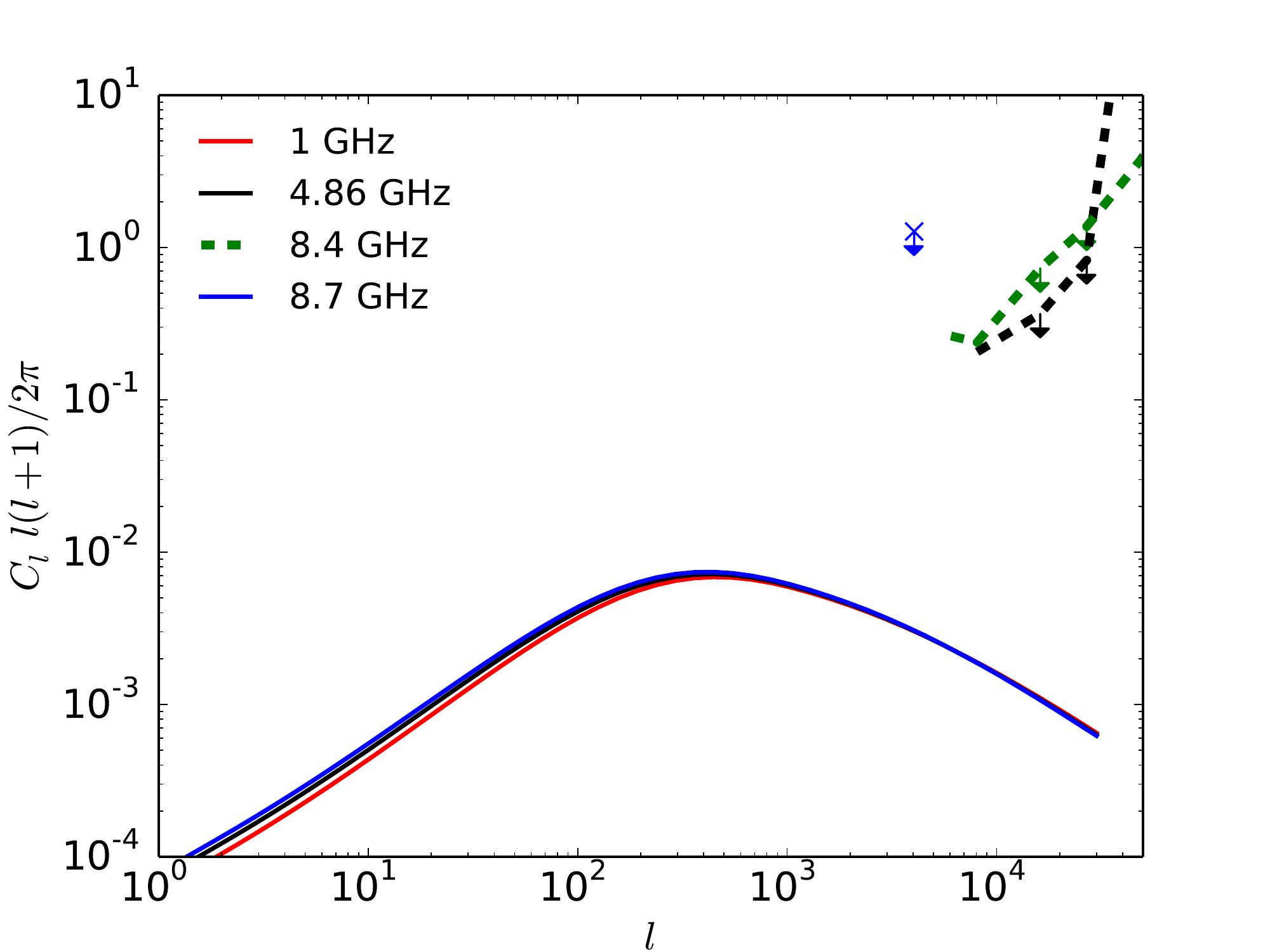,width=0.45\textwidth}   
\caption{\label{fig:50GeV} The contribution to the isotropic radio background from dark matter particles with a mass of $50\,\rm GeV$ annihilating to $b\bar{b}$, as motivated by models of the galactic center excess~\citep{Calore:2014xka} (Case I in Table~\ref{table:input}). The annihilation cross section has been chosen to fit the data.  We adopt a substructure distribution that is calibrated to the Via Lactea simulation \citep{2010PhRvD..81d3532K}, but extrapolated to a maximum radius $r_{\rm sub}= 8\,r_{\rm vir}$. The magnetic field extending to $r_{\rm sub}$ is assumed to be have an strength of at least $B_{\rm sub}^* = 8\,\mu$G for a $10^{14}\,M_\odot$ halo, and scale as $M^{0.3}$ for larger or smaller halos. In the top panel, the synchrotron emission is shown to be comparable to the temperature excess reported by the ARCADE-2 collaboration \citep{arcade_interpretation} with $\chi^2=72.64$. In the bottom panel,  the expected anisotropy of the dark matter  signal at 1 GHz  (shown in three redshift bins), 4.86 GHz and 8.7 GHz are compared with the observational limits from ATCA and VLA at 4.86 GHz (black dashed), 8.4 GHz (green dashed) and 8.7 GHz (blue data point) \citep{formalont_extragalactic_anisotropy_4.85GHz, partridge_extragalactic_anisotropy_8.4GHz, subrahmanyan_extragalactic_anisotropy_8.7GHz}. }
\end{figure}

Case I consists of a 50~GeV dark matter particle annihilating to $b\bar{b}$ final states, which was characterized as a reasonable fit to the $\gamma$-ray excess in the inner galaxy~\citep{Calore:2014xka}. In Figure~\ref{fig:50GeV}, we calculate the radio temperature produced by this model as a function of frequency (top) and the anisotropy of the dark matter emission, compared to constraints from \citep{holder_anisotropy_of_arcade}. We find that the soft-spectrum from annihilations to $b\bar{b}$ final states is a relatively poor fit to the spectrum of the excess at frequencies above ~$\sim$300~MHz. The model thus provides a poor fit to the ARCADE-2 data ($\chi^2$~=~72.64), indicating that a harder leptonic component is necessary in order to reproduce the spectrum of the excess. Additionally, we note that annihilations to heavy quark final states convert a negligible fraction of their annihilation energy to e$^+$e$^-$ pairs, and a large fraction of their annihilation to $\gamma$-rays. Thus, this model requires both extremely large magnetic field and substructure contributions ($r_{\rm sub}= 8\,r_{\rm vir}$, $B_{\rm sub}^* = 8\,\mu$G), as well as an annihilation cross-section of 3.0~$\times$~10$^{-26}$~cm$^{3}$s$^{-1}$, which exceeds the best fit cross-section of the galactic center excess by about a factor of two, and is currently in moderate tension with the most recent constraints from dwarf-spheroidal galaxies~\citep{GeringerSameth:2011iw, Ackermann:2013yva, Geringer-Sameth:2014qqa, Anderson_Symposium}. 

Before moving on, however, it is worth noting the effect of a relatively soft temperature spectrum on the radio anisotropy limits. Note that the limits from \citet{holder_anisotropy_of_arcade} are calculated based on the total antennae temperature at 4.86, 8.4 and 8.7 GHz and the $b\bar{b}$ emission produces a negligible fraction of the total antennae temperature at these frequencies. Thus, the anisotropy of this model can be extremely large without overproducing the total radio anisotropy. Note that in each case the specific limits on the anisotropy of the dark matter synchrotron emission are recomputed based on the total power of the ARCADE-2 excess at each test frequency. 

\begin{figure}
\centering
\epsfig{file=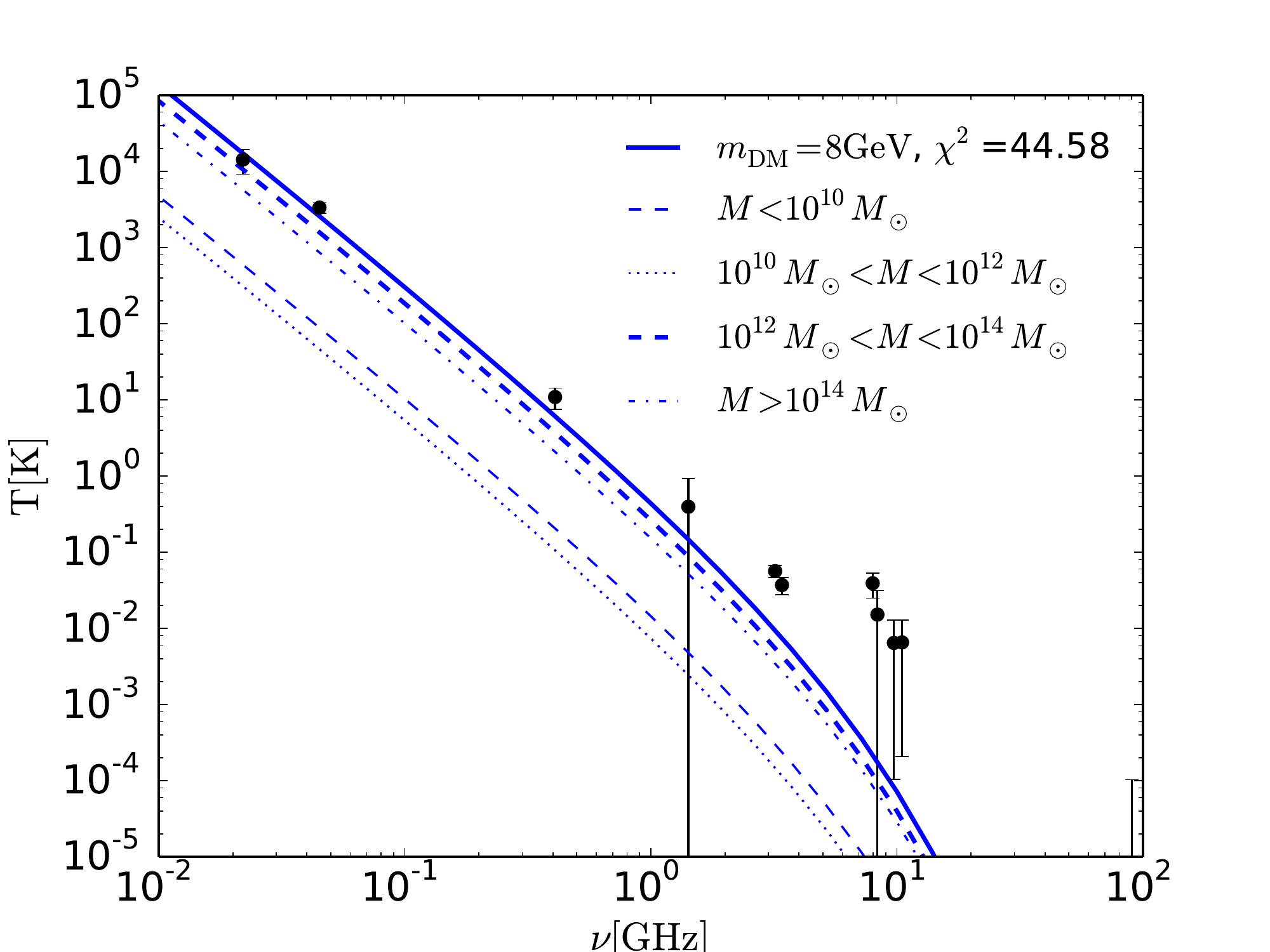,width=0.45\textwidth}   
\epsfig{file=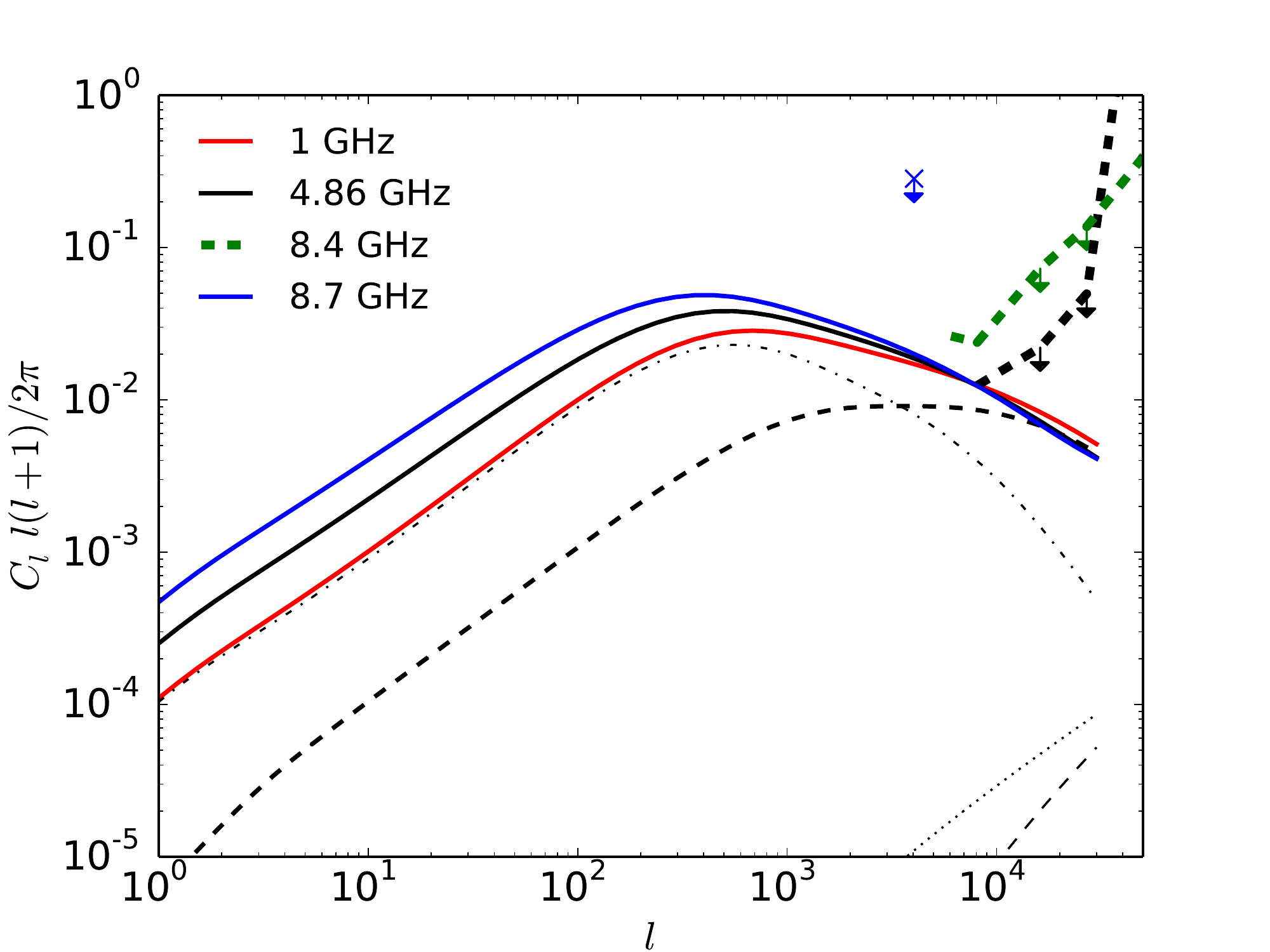,width=0.45\textwidth}   
\caption{\label{fig:8GeV} As in Figure~\ref{fig:50GeV} but for Case II in Table~\ref{table:input}, where the dark matter is assumed to have a mass of $8\,\rm GeV$ and annihilate democratically to leptonic final states (33\% to e$^+$e$^-$, 33\% to $\mu^+\mu^-$ and 33\% $\tau^+\tau^-$)\citep{hooper_linden_gc, linden_filamentary_arcs, Lacroix:2014eea}. In this case, the dark matter substructure is required to extend only to $r_{\rm sub} = 4\,r_{\rm vir}$ and the lower limit on the magnetic field is only $B_{\rm sub}^*=4\,\mu$G. This model provides a $\chi^2=44.58$ fit to the excess data while remaining consistent with the upper limits of the anisotropy. In this figure we additionally include the sub-contributions to the total ARCADE-2 temperature stemming from dark matter halos of various masses. In this figure we additionally show the contributions from four mass groups: $M<10^{10}\,M_\odot$ (thin dashed), $10^{10}M_\odot<M<10^{12}\,M_\odot$ (dotted), $10^{12}M_\odot<M<10^{14}M_\odot$ (dash-dot), and $M>10^{14}M_\odot$ (thick dashed). }
\end{figure}

Case II assumes the annihilation of an 8~GeV dark matter particle democratically to leptonic final states (i.e. 33\% to e$^+$e$^-$, 33\% to $\mu^+\mu^-$, and 33\% to $\tau^+\tau^-$). This model produces a harder input electron spectrum, greatly improving its fit to the ARCADE-2 data. In Figure~\ref{fig:8GeV}, we again plot the dark matter fit to the ARCADE-2 spectrum (top) and the anisotropy (bottom). In this case we find that the spectrum of the radio emission fits observations well up to a frequency of $\sim$4~GHz, when it begins to fall below the best fit observations. We note that the statistical fit to the ARCADE-2 data is still poor, ($\chi^2$~=~44.58). However, we note that this statistical fit includes only statistical errors in the ARCADE-2 excess, while large systematic errors due to foreground subtraction, and the existence of undetected point sources, may also be present~\citep{arcade_measurement, arcade_interpretation}. We thus consider this to be a reasonable fit to the ARCADE-2 data. Due to the diminished intensity at high frequencies, the signal is consistent with current constraints on the ARCADE-2 anisotropy, with the strongest constraints stemming from relatively low-frequency observations at 4.86~GHz. Moreover, due to the efficient conversion of the dark matter annihilation power into e$^+$e$^-$ pairs, the limits on the dark matter substructure contribution and the cluster magnetic field are significantly reduced to values of $r_{\rm sub}$~=~4~$r_{\rm vir}$ and $B_{\rm sub}^*$~=~4~$\mu$G. 

In Figure~\ref{fig:8GeV}, we additionally break down the total ARCADE-2 temperature spectrum into contributions based on the size of the galactic halo in question. We find that the total synchrotron emission is mostly split  between halos with masses between 10$^{12}$--10$^{14}$~M$_\odot$, and halos with masses $>$10$^{14}$~M$_\odot$, with negligible contributions from smaller halos.  This indicates that the majority of the ARCADE-2 emission is produced by structures significantly more massive than the Milky Way. 

\begin{figure}
\centering
\epsfig{file=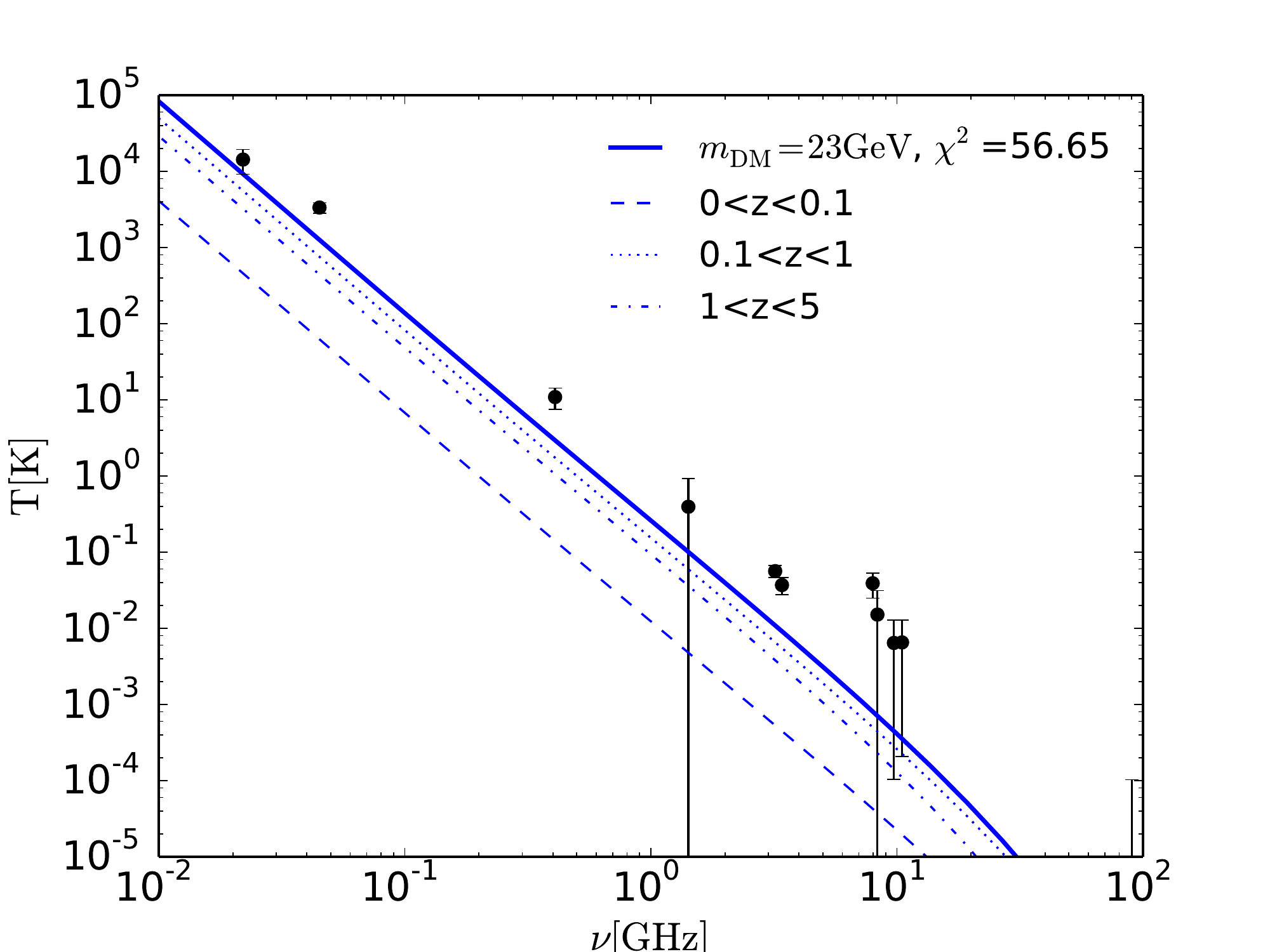,width=0.45\textwidth}   
\epsfig{file=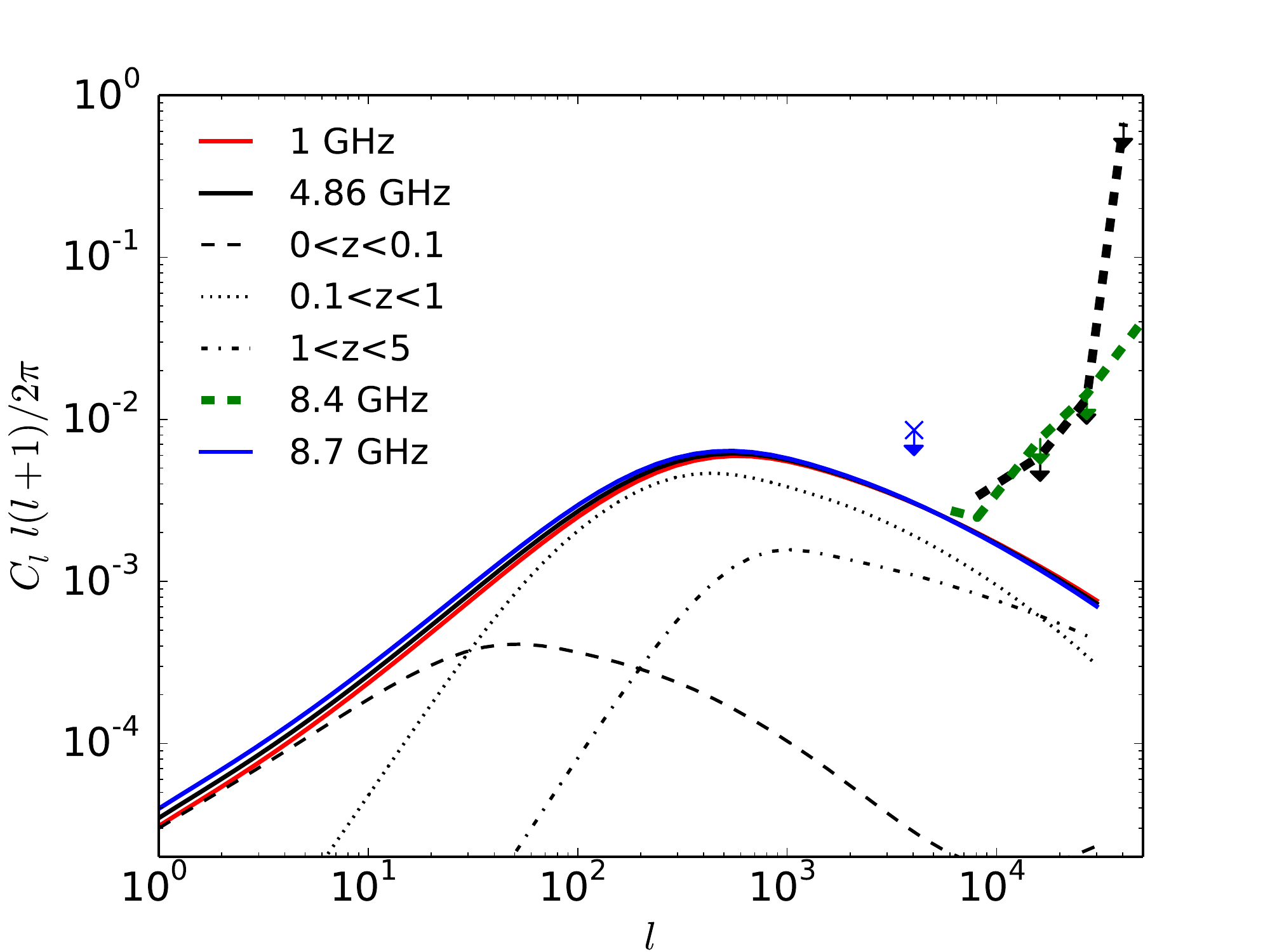,width=0.45\textwidth}   
\caption{\label{fig:23GeV} As in Fig.~\ref{fig:50GeV}, but for Case III in Table~\ref{table:input}. Specifically, we assume a dark matter particle with a mass of 23~GeV annihilating to all kinematically accessible standard model final states with a coupling constant proportionally to the square of the charge of the standard model particle. For this model we extend the substructure contribution out to $r_{\rm sub} = 8\,r_{\rm vir}$. The lower limit of the magnetic field  that extend to $r_{\rm sub}$ is taken to be $B_{\rm sub}^*=8\,\mu$G. This model provides a poor  fit  ($\chi^2=56.65$) to the excess data due to the soft intensity spectrum. As a consequence, the   upper limits on the anisotropy of this small portion of signal at $5-9\,\rm GHz$ are high. In this figure we additionally show the contributions to the total temperature from structure at distances 0~$<$~z~$<$~0.1, (dashed), 0.1~$<$~z~$<$~1.0 (dotted), and 1~$<$~z~$<$~5, (dash-dot). }
\end{figure}

Finally, in Case III, we assume a dark matter model that couples to all charged standard model particles, with a coupling constant proportional to the square of the particle charge. Such a coupling is well motivated in models where dark matter annihilates through a dark photon that kinetically mixes with the standard model photon. In Figure~\ref{fig:23GeV} we take a dark matter mass of 23~GeV, and find a reasonable match to both the temperature ($\chi^2$~=~56.65) and the anisotropy of the ARCADE-2 excess. Due to the slightly less efficient conversion of annihilation power into e$^+$e$^-$ pairs, a larger magnetic field is necessary than in case II, and we take values of $r_{\rm sub}$~=~8~$r_{\rm vir}$ and $B_{\rm sub}^*$~=~8~$\mu$G. 

In Figure~\ref{fig:23GeV} we additionally decompose the total contribution to both the synchrotron temperature and the anisotropy into different redshift bins, assuming a frequency of 1~GHz for the anisotropy calculation. We show results for $z<0.1$ (dashed), $0.1<z<1$ (dotted), and $1<z<5$ (dot-dashed). We find that the total antennae temperature is produced mostly by emission in the region $0.1<z<1$, with an important contribution from sources at $z<0.1$. Emission from redshifts z~$>$~1 provide less than 5\% of the total synchrotron power. This is primarily due to the fact that the CMB energy density increases rapidly with redshift and provides the dominant mechanism for electron energy loss at high redshifts. The redshift dependence of the anisotropy is more straightforward to understand, nearby sources produce emission on relatively large patches of the sky, and thus contribute anisotropies primarily at low-multiples. High redshift sources contribute anisotropies at much smaller angular scales. Note that the anisotropies at high multipole continue to be dominated by objects at moderate redshift, rather than high redshift sources, due to the much larger intensity of sources in the region $0.1<z<1$.

\begin{figure}
\centering
\epsfig{file=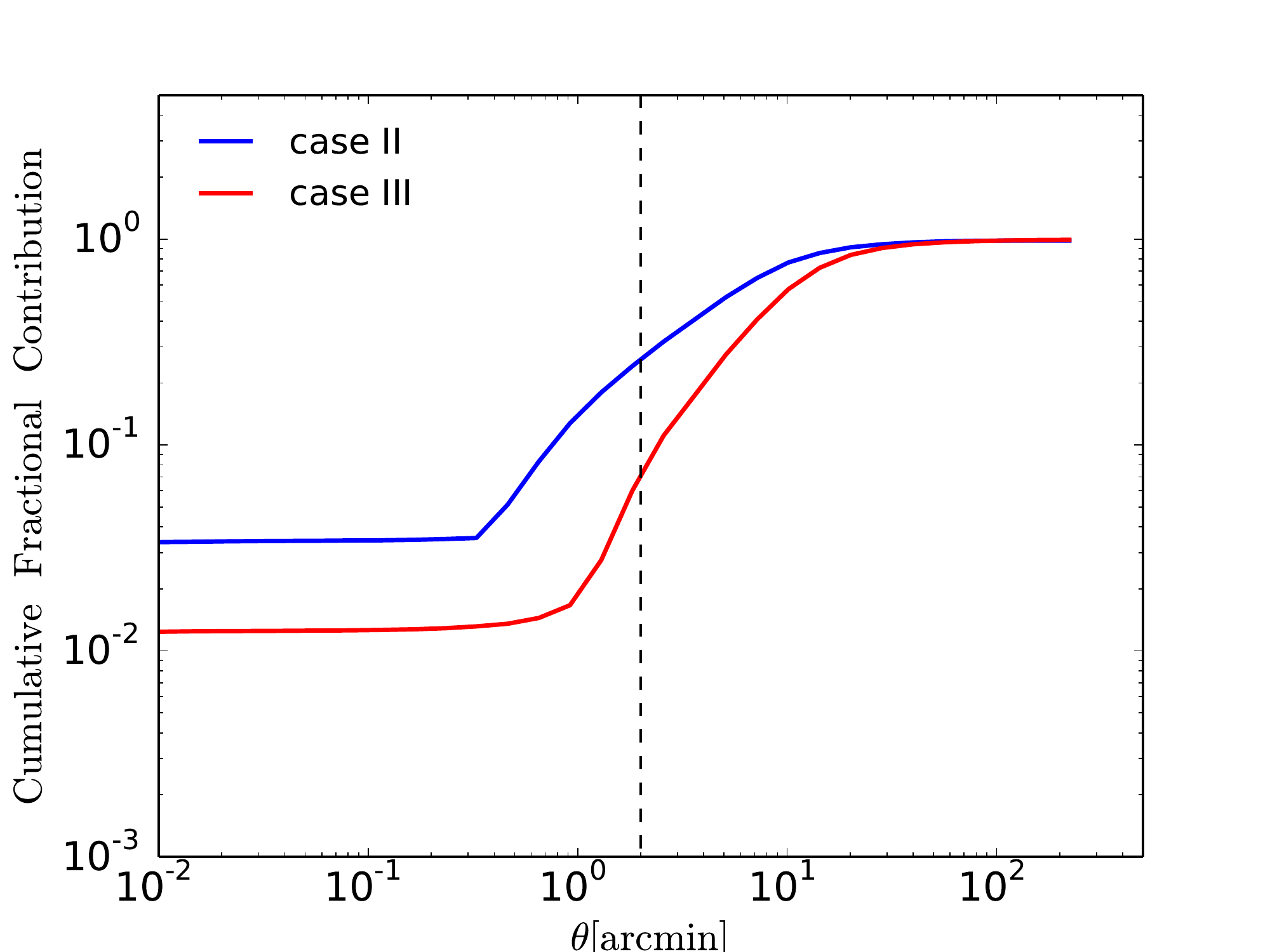,width=0.45\textwidth}   
\caption{\label{fig:intensity_fraction} The cumulative fractional contribution from dark matter structures of different sizes to the intensity of the ARCADE-2 excess. The angular size of a source is calculated as the radius that encloses half of the total synchrotron flux of the main halo. In Case II, with relatively small magnetic fields and maximum substructure contributions, 81\% of the emission is contributed by structures larger than 2'. In Case III, these large scale structures produce 96\% of the total emission. Since these structures are not resolved in the analysis of \citet{Vernstrom:2014uda}, it may be possible for dark matter to produce the intensity of the ARCADE-2 signal without overproducing the number of observed radio sources.}
\end{figure}

In Figure~\ref{fig:intensity_fraction} we show the cumulative fractional intensity of the ARCADE-2 excess provided by structures with various intrinsic angular scales. In particular, we determine the angular size of a source  as the radius that encloses half of its total synchrotron flux. Interestingly, we find that in both Case II and Case III, the intensity of the ARCADE-2 emission is dominated by structures larger than the 2' cutoff employed by \citet{Vernstrom:2014uda}. This result is specific to dark matter models, where the majority of the emission is produced by cluster-scaled objects, and the magnetic field and substructure contributions can extend farther than the virial radius. As a note, the changes in the shape of the cumulative fractions at around 0.3 arcminute for Case II and 1 arcminute for Case III are produced by the change in the magnetic field models for galaxy-sized halos and cluster-sized halos (see Eqn.~\ref{eqn:B_cluster} and \ref{eqn:B_galaxy} for details).  As a result of their condensed magnetic field structure, the angular sizes of halos with mass lower than $10^{12}\,M_\odot$ are mostly less than 0.01 arcminute.

\begin{figure*}
\centering
\epsfig{file=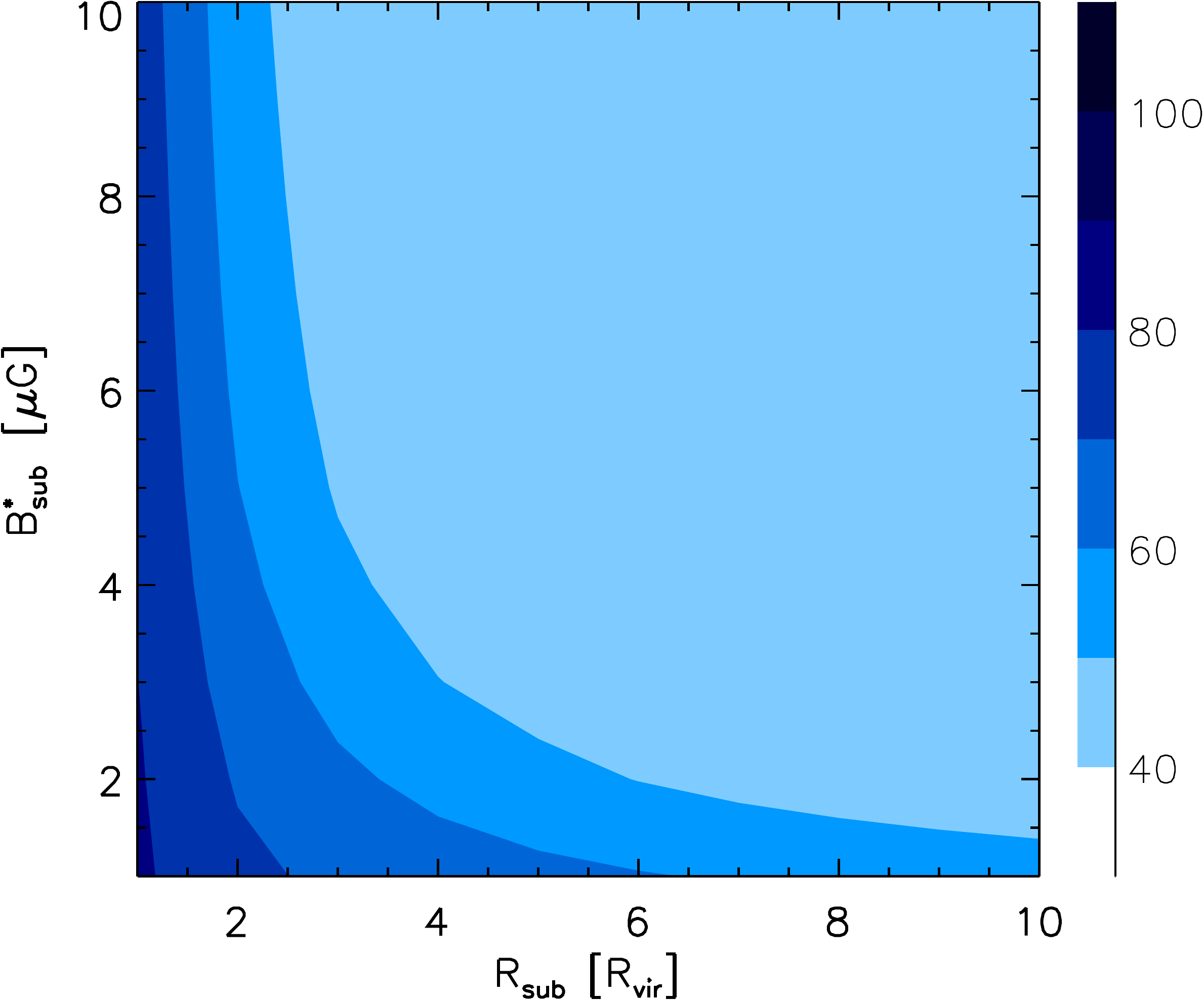,width=0.45\textwidth}  
\epsfig{file=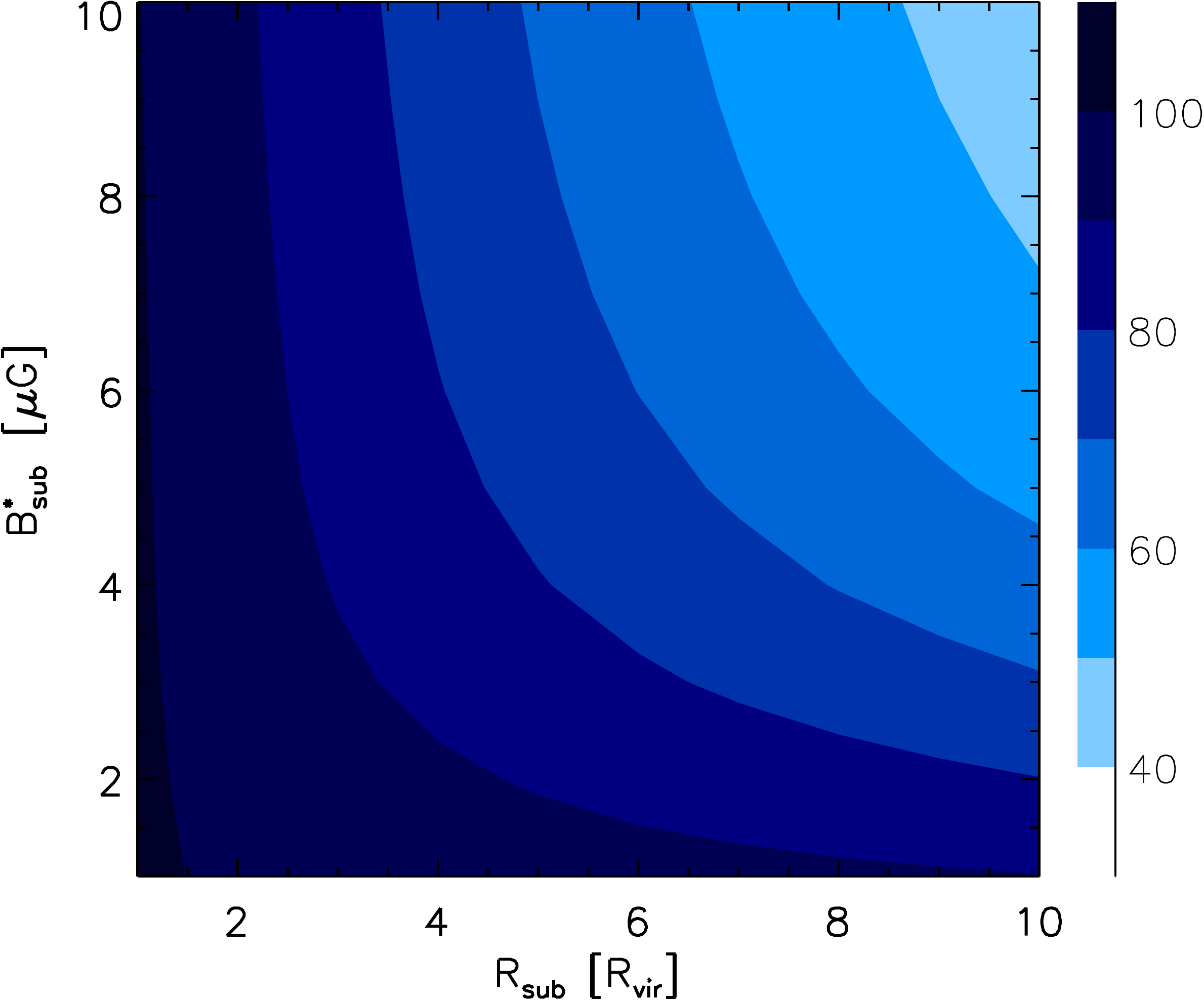,width=0.45\textwidth}  
\caption{\label{fig:parameter_space} The allowable parameter space for the termination of the dark matter substructure contribution $r_{\rm sub}$ and the minimum magnetic field strength in the cluster $B_{\rm sub}^*$ for dark matter with the parameters of Case II (left) and Case III (Right).  The contours refer to the best $\chi^2$ fit of the dark matter induced synchrotron signal to the spectrum and intensity of the ARCADE-2 excess, after allowing the annihilation cross-section to float, but removing any model that overproduces the anisotropy observed by \citep{holder_anisotropy_of_arcade}.}
\end{figure*}

As discussed in Section~\ref{subsec:B}, large uncertainties exist in the modeling of dark matter substructure and cluster magnetic fields. In order to understand the dependence of our results on these two factors, we scan the parameter space of $r_{\rm sub}$ and $B_{\rm sub}^*$ for our case II and case III dark matter models. In Figure~\ref{fig:parameter_space}, we show the resulting $\chi^2$ fits to the ARCADE-2 temperature spectrum. We note that in each simulation the dark matter annihilation rate is allowed to float in order to achieve the best fit to the ARCADE-2 excess, but a model is disregarded if it exceeds the anisotropy constraints from \citep{holder_anisotropy_of_arcade}.  For the $8\,\rm GeV$ dark matter model, the best fit $\chi^2\approx 40$ can always be realized when $r_{\rm sub}\gsim4\,r_{\rm vir}$ and $B^*_{\rm sub}\gsim 4\,\mu$G. Strong anisotropies would be produced if dark matter substructure stops contributing to the total annihilation rate at a distance smaller than $2\,r_{\rm vir}$. In order for the emission from this region to remain relevant, the model also demands that the minimum magnetic field exceed 4~$\mu$G for a $10^{14}\,M_\odot$ halo (corresponding to $1\,\mu$G for a $10^{12}\,M_\odot$ halo). For our case III model, the smaller e$^+$e$^-$ power necessitates larger substructure contributions and magnetic field strengths, finding reasonable fits  only if $r_{\rm sub}\gsim 8\,r_{\rm vir}$ and $B_{\rm sub}^*\gsim 8\,\mu$G.

\subsection{Multimessenger Constraints}

In the previous subsection, we have shown that our case II and case III models provide reasonable fits to the spectrum and isotropy of the ARCADE-2 excess. However, we note that both of these models require considerable annihilation rates to e$^+$e$^-$ pairs in order to produce the hard spectrum of the observed signal. These models are strongly constrained by local measurements of the positron fraction by AMS-02. In particular, our case II and case III models feature annihilation cross-sections directly to e$^+$e$^-$ final states of 2.8~$\times$~10$^{-27}$~cm$^{3}$s$^{-1}$ and 1.1~$\times$~10$^{-27}$~cm$^{3}$s$^{-1}$ respectively, both of which are currently ruled out by AMS-02 observations~\citep{Bergstrom:2013jra}. Since hard electron populations are necessary to explain the spectrum of the ARCADE-2 excess, it was recently argued that AMS-02 observations may rule out any dark matter interpretation of the signal~\citep{Fairbairn:2014nga}. 

However, we note that this constraint can be lessened somewhat in scenarios where the dark matter annihilation event does not directly produce leptonic pairs, but instead annihilates through a dark force mediator with a mass m$_\phi$~$\ll$~m$_\chi$, which itself decays to standard model pairs. In this case, the same fraction of the dark matter annihilation power is still transferred into e$^+$e$^-$, but the spectrum is smeared out by the kinematics of the light force mediator. It is easy to estimate the effect of a light mediator on the AMS-02 constraints obtained by \citep{Bergstrom:2013jra} by noting that annihilations to $\mu^+\mu^-$ approximate annihilations through a light mediator so long as m$_\chi$~$\gg$~m$_\mu$ and the internal bremsstrahlung of muons is neglected. However, the muon decay also produces relativistic neutrinos which carry away 2/3 of the total muon energy. Thus, the constraints from annihilations through a light mediator are approximately a factor of 3 stronger than annihilations through $\mu^+\mu^-$, and approximately a factor of 3 above the constraints from direct annihilations to e$^+$e$^-$. Taking the constraints of \citep{Bergstrom:2013jra} and re-normalizing the local density from a value of 0.4~GeV~cm$^{-3}$ to 0.3~GeV~cm$^{-3}$, we obtain very approximate constraints on the annihilation of dark matter through a light mediator to e$^+$e$^-$ pairs of  5~$\times$~10$^{-28}$~cm$^{3}$s$^{-1}$ and  1.5~$\times$~10$^{-27}$~cm$^{3}$s$^{-1}$ for dark matter masses of 8~GeV and 23~GeV respectively. While this appears to still create tension with case II, case III remains consistent with AMS-02 constraints. Furthermore, we note that the existence of a light mediator in charge-coupled models (such as case III) is particularly well motivated. In this class of models, the role of m$_\phi$ is typically filled by a dark photon which kinetically mixes with the standard model photon. In the case that the mass m$_\phi$ lies above twice the $\tau$ mass, the scenario above can be replicated. We note that the generic properties of charge-couple scenarios remain consistent even for masses of m$_\phi$~$\sim$~1~GeV, where the dark matter is kinematically forbidden from annihilating to $b\bar{b}$ and $\tau^+\tau^-$ final states.

Finally, we note an additional constraint from cosmic microwave background (CMB) experiments such as Planck. Annihilations to e$^+$e$^-$ produce reheating in the early universe which alters the properties of recombination~\citep{Slatyer:2009yq}. These constraints are particularly difficult to avoid in ARCADE-2 models, due to the fact that every electron which contributes to the synchrotron signal from the ARCADE-2 excess also contributes to the reheating of the early universe. Predictions for Planck sensitivity indicate that our models are expected to lie on the precipice of Planck detection or constraint.

\section{Enhancement from Alfv\'en Reacceleration}\label{sec:Alfven}

In the standard scenario, e$^+$e$^-$ produced via dark matter annihilation cool via synchrotron and inverse-Compton scattering. Upon losing the majority of their energy to these processes, the electrons remain non-relativistic indefinitely. However, it is possible for these e$^+$e$^-$ to be reaccelerated by turbulence in the ambient medium, allowing these electrons to again produce synchrotron emission. This offers a separate method for substantially enhancing the dark matter synchrotron flux, in addition to dark matter substructure and strong magnetic fields. Interestingly, the Alfv\'enic reacceleration of low-energy e$^+$e$^-$ is particularly effective in the outer regions of galaxy clusters, where the gas densities are negligible, meaning that electrons lose energy primarily via inverse Compton scattering and synchrotron radiation, and can survive in the intra-Cluster medium for extremely long periods without annihilating or recombining with ionized gas. 

Observations of radio halos indicate that cluster mergers can produce energetic Alfv\'en waves that channel energy from the cluster collision into the acceleration of charged particles in the intra-Cluster medium~\citep{Ferrari:2008jr, 2008Natur.455..944B}. Through interaction with Alfv\'en waves, electrons can be accelerated to energies as high as $\sim$~5~GeV, an energy which is compatible with fits to the ARCADE-2 excess \citep{2004MNRAS.350.1174B, 2005MNRAS.363.1173B}. In some models, cluster shocks are even capable of providing enough energy to explain the entirety of the ARCADE-2 signal through the reacceleration of thermal electrons. Additionally, cluster shocks provide the larger magnetic fields necessary in order to efficiently convert electron energy into synchrotron radiation~\citep{2006AJ....131.2900C, 2010Sci...330..347V, 2005AN....326..414B}. However, these models depend sensitively on the properties of cluster shocks in both major and minor mergers. The exploration of this parameter space lies outside the scope of this text, but will be investigated in future work~\citep{fang_linden_15}.

In this paper, we estimate the enhancement to the dark matter annihilation luminosity from Alfv\'en reacceleration of the leptons produced in the initial annihilation event. We follow the qausi-static calculation of the injected power in Alfv\'en waves from \citet{2004MNRAS.350.1174B}:
\begin{equation}
P_A(r) = P_A(r=0)\left(\frac{n_{\rm th}(r)}{n_{\rm th}(r=0)}\right)^{5/6}
\end{equation} 
where $P_A(r=0)$ is the injection rate of the Alfv\'en waves in the center of the cluster, which is set to be $10^{-29}\,\rm erg~s^{-1}~cm^{-3}$  \citep{2004MNRAS.350.1174B}. $n_{\rm th}$ is the radial density profile of the thermal gas in the  intercluster medium, in the form  \citep{1976A&A....49..137C}:
\begin{equation}
n_{\rm th} = n_{\rm th}(r=0)\left[1+\left(\frac{r}{r_c}\right)^2\right]^{-3\beta/2}
\end{equation}
where $r_c$ is the core radius and $\beta$ is the profile index, for which we adopt $\beta=0.6$. The Alfv\'en modes are estimated to last for  $\tau_A \sim 0.5-0.7$ Gyr for a Coma-like cluster \citep{2004MNRAS.350.1174B}. We further assume that $\tau_A$ scales as $M^{0.3}$ for halos of different masses. Noting that Alfv\'en waves provide a reaccelerated electron spectrum that is highly peaked at an energy of $\sim$5~GeV~\citep{2004MNRAS.350.1174B, 2005MNRAS.363.1173B}, similar to our dark matter models, in this work we do not consider changes to the electron spectrum due to  Alfv\'en reacceleration (essentially assuming that Alfv\'en waves maintain electrons at their steady state energy distribution). 

In our simplified model, the impact of Alfv\'enic reacceleration on the effective dark matter profile is taken into account by a modification to the effective dark matter annihilation rate $S_{\rm eff}$ (as defined in Eqn.~\ref{eqn:Seff}), which compares the Alfv\'enic power to the annihilation power:
\begin{equation}
S_{\rm eff}^A = S_{\rm eff}\,\,\left(\frac{P_A\tau_A\eta_A}{(\rho_{\rm DM}^2\langle \sigma v\rangle/2m_{\rm DM}^2)\,E_e\,\tau_{\rm cluster}}+1\right)
\end{equation}
Here $\eta_A$ is the efficiency of the Alfv\'en waves in reaccelerating the electrons.  We take $\eta\approx 0.1$, considering that despite the fact that most of the wave energy would first be converted into cosmic ray protons, it would  finally be channeled into electrons and positrons through the inelastic interactions of protons with the ambient gas \citep{2005MNRAS.363.1173B}.  The value $\tau_{\rm cluster}\approx 10\,\rm~Gyr$ is the age of the galaxies and clusters, and $E_e\sim m_{\rm DM}/2$ is the typical energy of electrons from the dark matter annihilation. 

\begin{figure}
\centering
\epsfig{file=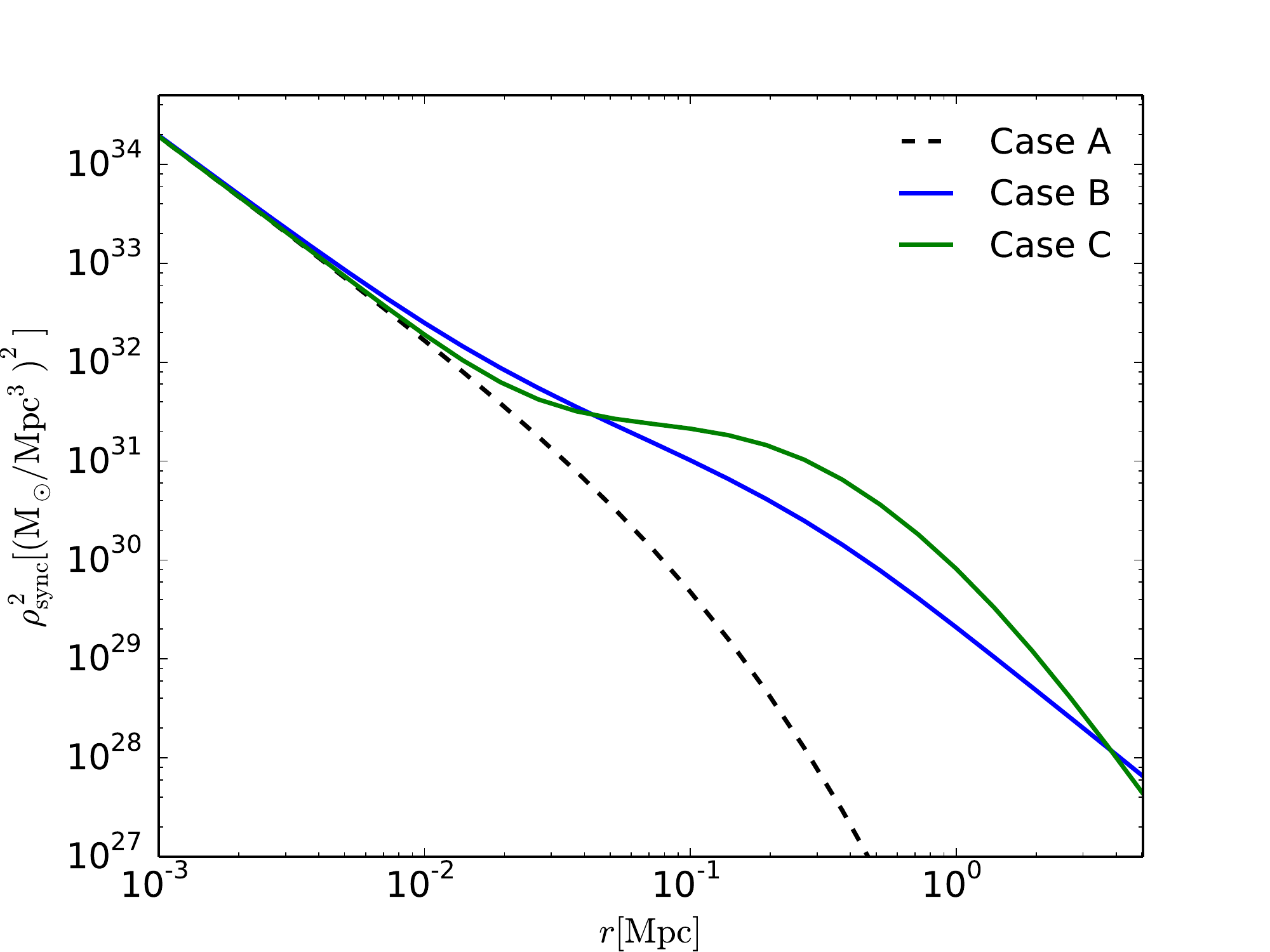,width=0.45\textwidth}  
\caption{\label{fig:M14_profiles} The radial distribution of the effective dark matter density squared $\rho^2_{\rm sync}$ that determines the production rate of synchrotron-emitting electrons for a $10^{14}\,M_\odot$ halo at $z=0.1$, for cases where only dark matter from the main halo contributes to the annihilation rate (case A, black dotted), where dark matter substructure contributes up to a maximum radius $r_{\rm sub}=4\,r_{\rm vir}$ with a minimum magnetic field strength $B_{\rm sub}=4\,\mu$G (case B, also case II in Sec.~\ref{sec:results}), and where there is no substructure or minimum magnetic field contribution, but electrons can be reaccelerated by Alfv\'en waves with a central power of $P_A(r=0)=10^{-29}\,\rm erg \,cm^{-3} s^{-1}$ and a duration of $\tau_A=0.5\,\rm Gyr$ (case C).}
\end{figure}

In Figure~\ref{fig:M14_profiles} we compare the effective annihilation profile from a 10$^{14}$~M$_\odot$ halo at $z=0.1$ for three different leptonic acceleration scenarios: case A shows the square of the effective dark matter density from the main halo after the fractional contribution of electrons to synchrotron radiation is taken to account, but without any substructure contribution, so that $\rho_{\rm sync}^2=\rho_{\rm NFW}^2(\rho_B/(\rho_B+\rho_{\rm CMB})$), case B follows case II of Section~\ref{sec:results} and includes the boost from substructure in an extended magnetic field with r$_{\mathrm{sub}}$~=~4~r$_{\mathrm{vir}}$ and  B$_{\mathrm{sub}}$~=~4~$\mu$G, case C ignores contributions from substructure and extended magnetic fields, but includes reacceleration of electrons from collisional shocks assuming a central power of $10^{-29}\,\rm erg~s^{-1}~cm^{-3}$ and an extension to $r_A\sim 4\,r_{\rm vir}$. Interestingly, we find that the effective annihilation profiles of cases B and C are nearly identical at distances far from the profile center, suggesting that reacceleration could efficiently recycle the relativistic electrons from dark matter annihilation and relax the large substructure boost factors and magnetic fields employed in Section~\ref{sec:results}.

\begin{figure}
\centering
\epsfig{file=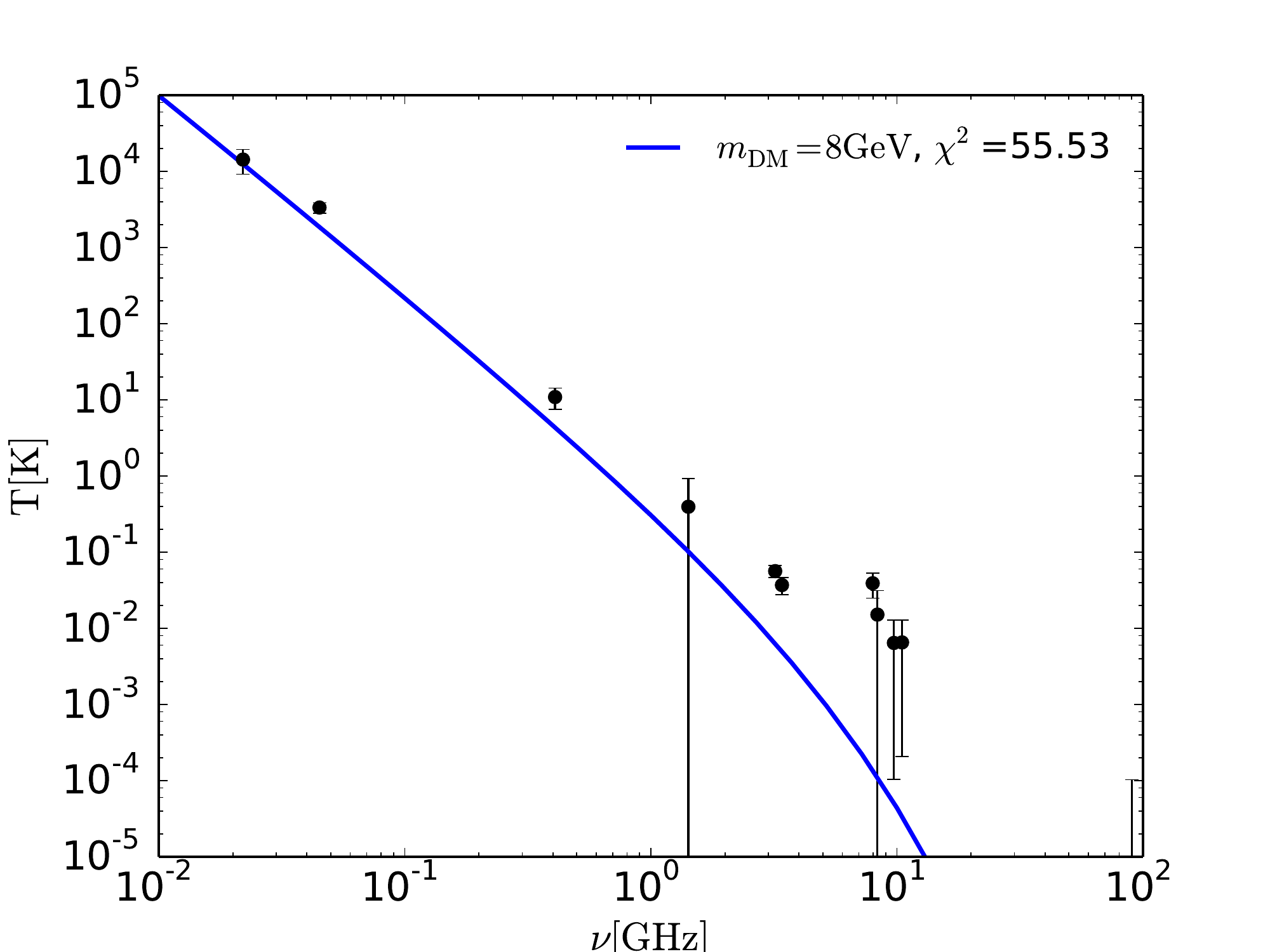,width=0.45\textwidth}  
\epsfig{file=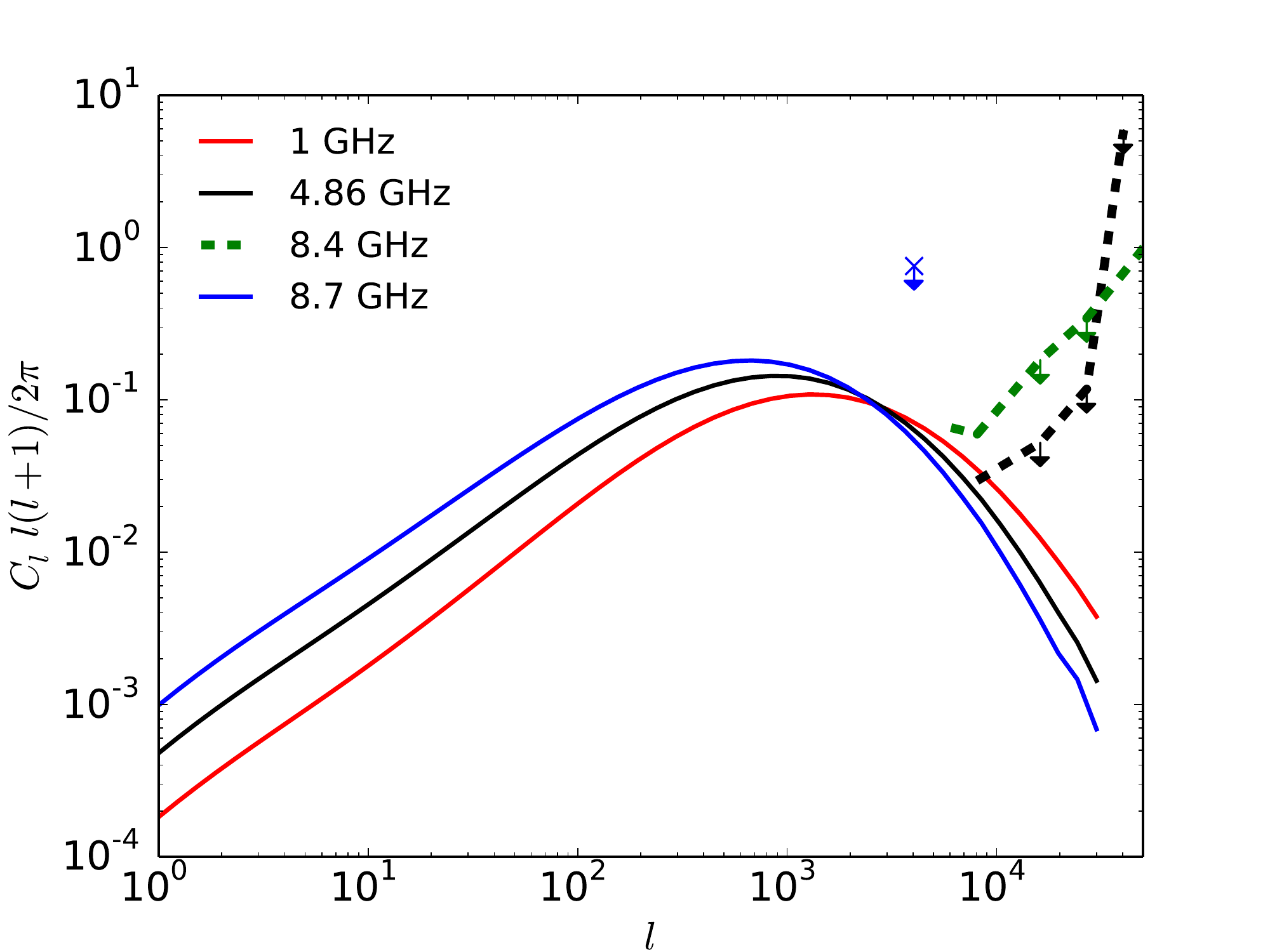,width=0.45\textwidth}  
\caption{\label{fig:Alfven} As in Fig.~\ref{fig:8GeV}, but for Case C as described  in Sec~\ref{sec:Alfven}. Specifically, the dark matter is assumed to have a mass of $8\,\rm GeV$ and annihilate democratically to leptonic final states (33\% to e$^+$e$^-$, 33\% to $\mu^+\mu^-$ and 33\% $\tau^+\tau^-$)\citep{hooper_linden_gc, linden_filamentary_arcs, Lacroix:2014eea}. In addition, the dark matter is assumed to follow a NFW profile, but the electrons from the annihilation are assumed to undergo  reacceleration   in the Alfv\'en waves powered by the cluster mergers \citep{2004MNRAS.350.1174B, 2005MNRAS.363.1173B}. This model provides a $\chi^2=55.53$ fit to the excess data while remaining consistent with the upper limits from the radio anisotropy studies.}
\end{figure}

In Fig.~\ref{fig:Alfven} we calculate the intensity and anisotropy of the synchrotron signal in Case C, for the case of an 8~GeV dark matter particle annihilating through the democratic channel with a cross section of $1.6\times10^{-27}\,\rm cm^3 s^{-1}$. Notice that in this case, neither the contribution from an extended substructure distribution nor an extended magnetic field is necessary ($B_{\rm sub}=0$). These results indicate that the Alfv\'enic reacceleration mechanism could effectively reproduce the intensity of the ARCADE-2 excess while maintaining consistency with anisotropy constraints. 

However, we note several important caveats to the currently presented calculation of Alfv\'enic reacceleration. In this paper, we have assumed that  Alfv\'en reacceleration occurs equally in all dark matter halos of equivalent mass, and have simply multiplied the luminosity from reaccelerated electrons in all halos by $\tau_A$ in order to capture the transient and stochastic nature of these events. This provides a correct calculation for the total intensity from Alfv\'en reacceleration, but underestimates the total anisotropy from this mechanism. Secondly, we have employed an approximate calculation of Alfv\'enic reacceleration and extended its effect to r$_{\mathrm{sub}}$~=~4~r$_{\mathrm{vir}}$. A full calculation of the effectiveness of reacceleration must include the direct calculation of the coupled magnetic-hydrodynamic equations. We will take into account these factors in future work. 

\section{Discussion and Conclusions}
\label{sec:conclusions}

We find that the isotropy of the ARCADE-2 excess strongly constrains dark matter interpretations of this signal. Models are forced into a relatively unsavory portion of the parameter space. Large magnetic fields must extend to several times the cluster virial radius, and the annihilation boost factors from dark matter substructure must be relatively large, compared to the recent constraints from~\citep{2014MNRAS.442.2271S}. Moreover, it is difficult to produce the significant e$^+$e$^-$ fluxes from dark matter annihilation in lieu of constraints from AMS-02 and Planck. 

However, it is worth noting that despite these issues, dark matter remains a relatively viable method for explaining this peculiar excess. Notably, dark matter models offer the unique ability to create significant emission on cluster scales, decreasing the anisotropy observed by radio telescopes. Astrophysical sources that trace large scale structure, such as AGN, radio galaxies, and star forming galaxies, are strongly ruled out by the analyses of \citet{holder_anisotropy_of_arcade} and \citet{Vernstrom:2014uda}. Competing models that have been produced to explain the observed isotropy of the ARCADE-2 data have also invoked novel physics, such as a phase transition at redshift z=5~\citep{Cline:2012hb}. 

Interestingly, mechanisms such as Alfv\'enic reacceleration may significantly weaken many of the constraints on dark matter models, by providing additional power in the intra cluster medium of the largest galaxy clusters. Due to uncertainties in the substructure boost factors and magnetic field strengths of galaxy clusters, as well as the energy density of Alfv\'en waves, it is difficult to precisely compute the interplay between the substructure and reacceleration scenarios without a precise understanding of each process. However, it is clear that both methods are independently powerful enough to produce the entirety of the ARCADE-2 excess, while remaining consistent with the observed anisotropy.  
 
Given the difficulty of fitting the ARCADE-2 data with any known astrophysical mechanism, dark matter annihilation remains a reasonable explanation for the ARCADE-2 excess due to the relatively high Bayesian prior on dark matter annihilation provided by the WIMP miracle. Viewed in this light, the tension between models of the ARCADE-2 excess, and constraints from AMS-02 offers the exciting promise for future confirmation of dark matter models of the ARCADE-2 excess. 

\acknowledgments
We would like to thank Benedikt Diemer, Gil Holder, Dan Hooper, Andrey Kravtsov, Yin Li, Angela Olinto and Stefano Profumo for fruitful discussions. TL is supported by the National Aeronautics and Space Administration through Einstein Postdoctoral Fellowship Award Number PF3-140110. KF acknowledge financial support from the NSF grant PHY-1068696 and  the NASA grant 11-APRA-0066 at  the University of Chicago, and the grant NSF PHY-1125897 at the Kavli Institute for Cosmological Physics. This work made use of computing resources and support provided by the Research Computing Center at the University of Chicago.

\bibliography{arcade} 

\begin{thebibliography}{74}
\expandafter\ifx\csname natexlab\endcsname\relax\def\natexlab#1{#1}\fi
\expandafter\ifx\csname bibnamefont\endcsname\relax
  \def\bibnamefont#1{#1}\fi
\expandafter\ifx\csname bibfnamefont\endcsname\relax
  \def\bibfnamefont#1{#1}\fi
\expandafter\ifx\csname citenamefont\endcsname\relax
  \def\citenamefont#1{#1}\fi
\expandafter\ifx\csname url\endcsname\relax
  \def\url#1{\texttt{#1}}\fi
\expandafter\ifx\csname urlprefix\endcsname\relax\def\urlprefix{URL }\fi
\providecommand{\bibinfo}[2]{#2}
\providecommand{\eprint}[2][]{\url{#2}}

\bibitem[{\citenamefont{{Seiffert} et~al.}(2009)\citenamefont{{Seiffert},
  {Fixsen}, {Kogut}, {Levin}, {Limon}, {Lubin}, {Mirel}, {Singal}, {Villela},
  {Wollack} et~al.}}]{arcade_interpretation}
\bibinfo{author}{\bibfnamefont{M.}~\bibnamefont{{Seiffert}}},
  \bibinfo{author}{\bibfnamefont{D.~J.} \bibnamefont{{Fixsen}}},
  \bibinfo{author}{\bibfnamefont{A.}~\bibnamefont{{Kogut}}},
  \bibinfo{author}{\bibfnamefont{S.~M.} \bibnamefont{{Levin}}},
  \bibinfo{author}{\bibfnamefont{M.}~\bibnamefont{{Limon}}},
  \bibinfo{author}{\bibfnamefont{P.~M.} \bibnamefont{{Lubin}}},
  \bibinfo{author}{\bibfnamefont{P.}~\bibnamefont{{Mirel}}},
  \bibinfo{author}{\bibfnamefont{J.}~\bibnamefont{{Singal}}},
  \bibinfo{author}{\bibfnamefont{T.}~\bibnamefont{{Villela}}},
  \bibinfo{author}{\bibfnamefont{E.}~\bibnamefont{{Wollack}}},
  \bibnamefont{et~al.}, \bibinfo{journal}{ArXiv e-prints}
  (\bibinfo{year}{2009}), \eprint{0901.0559}.

\bibitem[{\citenamefont{{Fixsen} et~al.}(2009)\citenamefont{{Fixsen}, {Kogut},
  {Levin}, {Limon}, {Lubin}, {Mirel}, {Seiffert}, {Singal}, {Wollack},
  {Villela} et~al.}}]{arcade_measurement}
\bibinfo{author}{\bibfnamefont{D.~J.} \bibnamefont{{Fixsen}}},
  \bibinfo{author}{\bibfnamefont{A.}~\bibnamefont{{Kogut}}},
  \bibinfo{author}{\bibfnamefont{S.}~\bibnamefont{{Levin}}},
  \bibinfo{author}{\bibfnamefont{M.}~\bibnamefont{{Limon}}},
  \bibinfo{author}{\bibfnamefont{P.}~\bibnamefont{{Lubin}}},
  \bibinfo{author}{\bibfnamefont{P.}~\bibnamefont{{Mirel}}},
  \bibinfo{author}{\bibfnamefont{M.}~\bibnamefont{{Seiffert}}},
  \bibinfo{author}{\bibfnamefont{J.}~\bibnamefont{{Singal}}},
  \bibinfo{author}{\bibfnamefont{E.}~\bibnamefont{{Wollack}}},
  \bibinfo{author}{\bibfnamefont{T.}~\bibnamefont{{Villela}}},
  \bibnamefont{et~al.}, \bibinfo{journal}{ArXiv e-prints}
  (\bibinfo{year}{2009}), \eprint{0901.0555}.

\bibitem[{\citenamefont{{Roger} et~al.}(1999)\citenamefont{{Roger}, {Costain},
  {Landecker}, and {Swerdlyk}}}]{roger_22MHz_excess}
\bibinfo{author}{\bibfnamefont{R.~S.} \bibnamefont{{Roger}}},
  \bibinfo{author}{\bibfnamefont{C.~H.} \bibnamefont{{Costain}}},
  \bibinfo{author}{\bibfnamefont{T.~L.} \bibnamefont{{Landecker}}},
  \bibnamefont{and} \bibinfo{author}{\bibfnamefont{C.~M.}
  \bibnamefont{{Swerdlyk}}}, \bibinfo{journal}{AAPS}
  \textbf{\bibinfo{volume}{137}}, \bibinfo{pages}{7} (\bibinfo{year}{1999}),
  \eprint{arXiv:astro-ph/9902213}.

\bibitem[{\citenamefont{{Guzm{\'a}n} et~al.}(2011)\citenamefont{{Guzm{\'a}n},
  {May}, {Alvarez}, and {Maeda}}}]{guzman_44Mhz_excess}
\bibinfo{author}{\bibfnamefont{A.~E.} \bibnamefont{{Guzm{\'a}n}}},
  \bibinfo{author}{\bibfnamefont{J.}~\bibnamefont{{May}}},
  \bibinfo{author}{\bibfnamefont{H.}~\bibnamefont{{Alvarez}}},
  \bibnamefont{and} \bibinfo{author}{\bibfnamefont{K.}~\bibnamefont{{Maeda}}},
  \bibinfo{journal}{AAP} \textbf{\bibinfo{volume}{525}}, \bibinfo{eid}{A138}
  (\bibinfo{year}{2011}), \eprint{1011.4298}.

\bibitem[{\citenamefont{{Haslam} et~al.}(1981)\citenamefont{{Haslam}, {Klein},
  {Salter}, {Stoffel}, {Wilson}, {Cleary}, {Cooke}, and
  {Thomasson}}}]{haslam_408Mhz_excess}
\bibinfo{author}{\bibfnamefont{C.~G.~T.} \bibnamefont{{Haslam}}},
  \bibinfo{author}{\bibfnamefont{U.}~\bibnamefont{{Klein}}},
  \bibinfo{author}{\bibfnamefont{C.~J.} \bibnamefont{{Salter}}},
  \bibinfo{author}{\bibfnamefont{H.}~\bibnamefont{{Stoffel}}},
  \bibinfo{author}{\bibfnamefont{W.~E.} \bibnamefont{{Wilson}}},
  \bibinfo{author}{\bibfnamefont{M.~N.} \bibnamefont{{Cleary}}},
  \bibinfo{author}{\bibfnamefont{D.~J.} \bibnamefont{{Cooke}}},
  \bibnamefont{and}
  \bibinfo{author}{\bibfnamefont{P.}~\bibnamefont{{Thomasson}}},
  \bibinfo{journal}{AAP} \textbf{\bibinfo{volume}{100}}, \bibinfo{pages}{209}
  (\bibinfo{year}{1981}).

\bibitem[{\citenamefont{{Reich} and {Reich}}(1986)}]{reich_1.4GHz_excess}
\bibinfo{author}{\bibfnamefont{P.}~\bibnamefont{{Reich}}} \bibnamefont{and}
  \bibinfo{author}{\bibfnamefont{W.}~\bibnamefont{{Reich}}},
  \bibinfo{journal}{AAPS} \textbf{\bibinfo{volume}{63}}, \bibinfo{pages}{205}
  (\bibinfo{year}{1986}).

\bibitem[{\citenamefont{{Kogut} et~al.}(2011)\citenamefont{{Kogut}, {Fixsen},
  {Levin}, {Limon}, {Lubin}, {Mirel}, {Seiffert}, {Singal}, {Villela},
  {Wollack} et~al.}}]{kogut_excess_not_galactic}
\bibinfo{author}{\bibfnamefont{A.}~\bibnamefont{{Kogut}}},
  \bibinfo{author}{\bibfnamefont{D.~J.} \bibnamefont{{Fixsen}}},
  \bibinfo{author}{\bibfnamefont{S.~M.} \bibnamefont{{Levin}}},
  \bibinfo{author}{\bibfnamefont{M.}~\bibnamefont{{Limon}}},
  \bibinfo{author}{\bibfnamefont{P.~M.} \bibnamefont{{Lubin}}},
  \bibinfo{author}{\bibfnamefont{P.}~\bibnamefont{{Mirel}}},
  \bibinfo{author}{\bibfnamefont{M.}~\bibnamefont{{Seiffert}}},
  \bibinfo{author}{\bibfnamefont{J.}~\bibnamefont{{Singal}}},
  \bibinfo{author}{\bibfnamefont{T.}~\bibnamefont{{Villela}}},
  \bibinfo{author}{\bibfnamefont{E.}~\bibnamefont{{Wollack}}},
  \bibnamefont{et~al.}, \bibinfo{journal}{ApJ} \textbf{\bibinfo{volume}{734}},
  \bibinfo{eid}{4} (\bibinfo{year}{2011}), \eprint{0901.0562}.

\bibitem[{\citenamefont{{Singal} et~al.}(2010)\citenamefont{{Singal},
  {Stawarz}, {Lawrence}, and
  {Petrosian}}}]{singal_not_extragalactic_baryonic_signals}
\bibinfo{author}{\bibfnamefont{J.}~\bibnamefont{{Singal}}},
  \bibinfo{author}{\bibfnamefont{{\L}.}~\bibnamefont{{Stawarz}}},
  \bibinfo{author}{\bibfnamefont{A.}~\bibnamefont{{Lawrence}}},
  \bibnamefont{and}
  \bibinfo{author}{\bibfnamefont{V.}~\bibnamefont{{Petrosian}}},
  \bibinfo{journal}{MNRAS} \textbf{\bibinfo{volume}{409}},
  \bibinfo{pages}{1172} (\bibinfo{year}{2010}), \eprint{0909.1997}.

\bibitem[{\citenamefont{{Gervasi} et~al.}(2008)\citenamefont{{Gervasi},
  {Tartari}, {Zannoni}, {Boella}, and
  {Sironi}}}]{gervasi_radio_luminosity_from_extrapolation_of_known_sources}
\bibinfo{author}{\bibfnamefont{M.}~\bibnamefont{{Gervasi}}},
  \bibinfo{author}{\bibfnamefont{A.}~\bibnamefont{{Tartari}}},
  \bibinfo{author}{\bibfnamefont{M.}~\bibnamefont{{Zannoni}}},
  \bibinfo{author}{\bibfnamefont{G.}~\bibnamefont{{Boella}}}, \bibnamefont{and}
  \bibinfo{author}{\bibfnamefont{G.}~\bibnamefont{{Sironi}}},
  \bibinfo{journal}{ApJ} \textbf{\bibinfo{volume}{682}}, \bibinfo{pages}{223}
  (\bibinfo{year}{2008}), \eprint{0803.4138}.

\bibitem[{\citenamefont{{Fornengo} et~al.}(2011)\citenamefont{{Fornengo},
  {Lineros}, {Regis}, and {Taoso}}}]{fornengo_arcade_excess_is_dm}
\bibinfo{author}{\bibfnamefont{N.}~\bibnamefont{{Fornengo}}},
  \bibinfo{author}{\bibfnamefont{R.}~\bibnamefont{{Lineros}}},
  \bibinfo{author}{\bibfnamefont{M.}~\bibnamefont{{Regis}}}, \bibnamefont{and}
  \bibinfo{author}{\bibfnamefont{M.}~\bibnamefont{{Taoso}}},
  \bibinfo{journal}{Physical Review Letters} \textbf{\bibinfo{volume}{107}},
  \bibinfo{eid}{271302} (\bibinfo{year}{2011}), \eprint{1108.0569}.

\bibitem[{\citenamefont{Goodenough and Hooper}(2009)}]{Goodenough:2009gk}
\bibinfo{author}{\bibfnamefont{L.}~\bibnamefont{Goodenough}} \bibnamefont{and}
  \bibinfo{author}{\bibfnamefont{D.}~\bibnamefont{Hooper}}
  (\bibinfo{year}{2009}), \eprint{0910.2998}.

\bibitem[{\citenamefont{{Hooper} and {Goodenough}}(2011)}]{hooper_goodenough}
\bibinfo{author}{\bibfnamefont{D.}~\bibnamefont{{Hooper}}} \bibnamefont{and}
  \bibinfo{author}{\bibfnamefont{L.}~\bibnamefont{{Goodenough}}},
  \bibinfo{journal}{Physics Letters B} \textbf{\bibinfo{volume}{697}},
  \bibinfo{pages}{412} (\bibinfo{year}{2011}), \eprint{1010.2752}.

\bibitem[{\citenamefont{{Hooper} and {Linden}}(2011)}]{hooper_linden_gc}
\bibinfo{author}{\bibfnamefont{D.}~\bibnamefont{{Hooper}}} \bibnamefont{and}
  \bibinfo{author}{\bibfnamefont{T.}~\bibnamefont{{Linden}}},
  \bibinfo{journal}{PRD} \textbf{\bibinfo{volume}{84}}, \bibinfo{eid}{123005}
  (\bibinfo{year}{2011}), \eprint{1110.0006}.

\bibitem[{\citenamefont{Abazajian and Kaplinghat}(2012)}]{Abazajian:2012pn}
\bibinfo{author}{\bibfnamefont{K.~N.} \bibnamefont{Abazajian}}
  \bibnamefont{and}
  \bibinfo{author}{\bibfnamefont{M.}~\bibnamefont{Kaplinghat}},
  \bibinfo{journal}{Phys.Rev.} \textbf{\bibinfo{volume}{D86}},
  \bibinfo{pages}{083511} (\bibinfo{year}{2012}), \eprint{1207.6047}.

\bibitem[{\citenamefont{Hooper and Slatyer}(2013)}]{Hooper:2013rwa}
\bibinfo{author}{\bibfnamefont{D.}~\bibnamefont{Hooper}} \bibnamefont{and}
  \bibinfo{author}{\bibfnamefont{T.~R.} \bibnamefont{Slatyer}},
  \bibinfo{journal}{Phys.Dark Univ.} \textbf{\bibinfo{volume}{2}},
  \bibinfo{pages}{118} (\bibinfo{year}{2013}), \eprint{1302.6589}.

\bibitem[{\citenamefont{Gordon and Macias}(2013)}]{Gordon:2013vta}
\bibinfo{author}{\bibfnamefont{C.}~\bibnamefont{Gordon}} \bibnamefont{and}
  \bibinfo{author}{\bibfnamefont{O.}~\bibnamefont{Macias}},
  \bibinfo{journal}{Phys.Rev.} \textbf{\bibinfo{volume}{D88}},
  \bibinfo{pages}{083521} (\bibinfo{year}{2013}), \eprint{1306.5725}.

\bibitem[{\citenamefont{Macias and Gordon}(2014)}]{Macias:2013vya}
\bibinfo{author}{\bibfnamefont{O.}~\bibnamefont{Macias}} \bibnamefont{and}
  \bibinfo{author}{\bibfnamefont{C.}~\bibnamefont{Gordon}},
  \bibinfo{journal}{Phys.Rev.} \textbf{\bibinfo{volume}{D89}},
  \bibinfo{pages}{063515} (\bibinfo{year}{2014}), \eprint{1312.6671}.

\bibitem[{\citenamefont{Abazajian
  et~al.}(2014{\natexlab{a}})\citenamefont{Abazajian, Canac, Horiuchi, and
  Kaplinghat}}]{Abazajian:2014fta}
\bibinfo{author}{\bibfnamefont{K.~N.} \bibnamefont{Abazajian}},
  \bibinfo{author}{\bibfnamefont{N.}~\bibnamefont{Canac}},
  \bibinfo{author}{\bibfnamefont{S.}~\bibnamefont{Horiuchi}}, \bibnamefont{and}
  \bibinfo{author}{\bibfnamefont{M.}~\bibnamefont{Kaplinghat}},
  \bibinfo{journal}{Phys.Rev.} \textbf{\bibinfo{volume}{D90}},
  \bibinfo{pages}{023526} (\bibinfo{year}{2014}{\natexlab{a}}),
  \eprint{1402.4090}.

\bibitem[{\citenamefont{Daylan et~al.}(2014)\citenamefont{Daylan, Finkbeiner,
  Hooper, Linden, Portillo et~al.}}]{Daylan:2014rsa}
\bibinfo{author}{\bibfnamefont{T.}~\bibnamefont{Daylan}},
  \bibinfo{author}{\bibfnamefont{D.~P.} \bibnamefont{Finkbeiner}},
  \bibinfo{author}{\bibfnamefont{D.}~\bibnamefont{Hooper}},
  \bibinfo{author}{\bibfnamefont{T.}~\bibnamefont{Linden}},
  \bibinfo{author}{\bibfnamefont{S.~K.~N.} \bibnamefont{Portillo}},
  \bibnamefont{et~al.} (\bibinfo{year}{2014}), \eprint{1402.6703}.

\bibitem[{\citenamefont{Abazajian
  et~al.}(2014{\natexlab{b}})\citenamefont{Abazajian, Canac, Horiuchi,
  Kaplinghat, and Kwa}}]{Abazajian:2014hsa}
\bibinfo{author}{\bibfnamefont{K.~N.} \bibnamefont{Abazajian}},
  \bibinfo{author}{\bibfnamefont{N.}~\bibnamefont{Canac}},
  \bibinfo{author}{\bibfnamefont{S.}~\bibnamefont{Horiuchi}},
  \bibinfo{author}{\bibfnamefont{M.}~\bibnamefont{Kaplinghat}},
  \bibnamefont{and} \bibinfo{author}{\bibfnamefont{A.}~\bibnamefont{Kwa}}
  (\bibinfo{year}{2014}{\natexlab{b}}), \eprint{1410.6168}.

\bibitem[{\citenamefont{Calore et~al.}(2014)\citenamefont{Calore, Cholis, and
  Weniger}}]{Calore:2014xka}
\bibinfo{author}{\bibfnamefont{F.}~\bibnamefont{Calore}},
  \bibinfo{author}{\bibfnamefont{I.}~\bibnamefont{Cholis}}, \bibnamefont{and}
  \bibinfo{author}{\bibfnamefont{C.}~\bibnamefont{Weniger}}
  (\bibinfo{year}{2014}), \eprint{1409.0042}.

\bibitem[{\citenamefont{{Holder}}(2012)}]{holder_anisotropy_of_arcade}
\bibinfo{author}{\bibfnamefont{G.}~\bibnamefont{{Holder}}},
  \bibinfo{journal}{ArXiv e-prints}  (\bibinfo{year}{2012}),
  \eprint{1207.0856}.

\bibitem[{\citenamefont{{Fomalont} et~al.}(1988)\citenamefont{{Fomalont},
  {Kellermann}, {Anderson}, {Weistrop}, {Wall}, {Windhorst}, and
  {Kristian}}}]{formalont_extragalactic_anisotropy_4.85GHz}
\bibinfo{author}{\bibfnamefont{E.~B.} \bibnamefont{{Fomalont}}},
  \bibinfo{author}{\bibfnamefont{K.~I.} \bibnamefont{{Kellermann}}},
  \bibinfo{author}{\bibfnamefont{M.~C.} \bibnamefont{{Anderson}}},
  \bibinfo{author}{\bibfnamefont{D.}~\bibnamefont{{Weistrop}}},
  \bibinfo{author}{\bibfnamefont{J.~V.} \bibnamefont{{Wall}}},
  \bibinfo{author}{\bibfnamefont{R.~A.} \bibnamefont{{Windhorst}}},
  \bibnamefont{and} \bibinfo{author}{\bibfnamefont{J.~A.}
  \bibnamefont{{Kristian}}}, \bibinfo{journal}{Astro. Journal}
  \textbf{\bibinfo{volume}{96}}, \bibinfo{pages}{1187} (\bibinfo{year}{1988}).

\bibitem[{\citenamefont{{Partridge} et~al.}(1997)\citenamefont{{Partridge},
  {Richards}, {Fomalont}, {Kellermann}, and
  {Windhorst}}}]{partridge_extragalactic_anisotropy_8.4GHz}
\bibinfo{author}{\bibfnamefont{R.~B.} \bibnamefont{{Partridge}}},
  \bibinfo{author}{\bibfnamefont{E.~A.} \bibnamefont{{Richards}}},
  \bibinfo{author}{\bibfnamefont{E.~B.} \bibnamefont{{Fomalont}}},
  \bibinfo{author}{\bibfnamefont{K.~I.} \bibnamefont{{Kellermann}}},
  \bibnamefont{and} \bibinfo{author}{\bibfnamefont{R.~A.}
  \bibnamefont{{Windhorst}}}, \bibinfo{journal}{ApJ}
  \textbf{\bibinfo{volume}{483}}, \bibinfo{pages}{38} (\bibinfo{year}{1997}).

\bibitem[{\citenamefont{{Subrahmanyan}
  et~al.}(2000)\citenamefont{{Subrahmanyan}, {Kesteven}, {Ekers}, {Sinclair},
  and {Silk}}}]{subrahmanyan_extragalactic_anisotropy_8.7GHz}
\bibinfo{author}{\bibfnamefont{R.}~\bibnamefont{{Subrahmanyan}}},
  \bibinfo{author}{\bibfnamefont{M.~J.} \bibnamefont{{Kesteven}}},
  \bibinfo{author}{\bibfnamefont{R.~D.} \bibnamefont{{Ekers}}},
  \bibinfo{author}{\bibfnamefont{M.}~\bibnamefont{{Sinclair}}},
  \bibnamefont{and} \bibinfo{author}{\bibfnamefont{J.}~\bibnamefont{{Silk}}},
  \bibinfo{journal}{MNRAS} \textbf{\bibinfo{volume}{315}}, \bibinfo{pages}{808}
  (\bibinfo{year}{2000}), \eprint{arXiv:astro-ph/0002467}.

\bibitem[{\citenamefont{Cline and Vincent}(2013)}]{Cline:2012hb}
\bibinfo{author}{\bibfnamefont{J.~M.} \bibnamefont{Cline}} \bibnamefont{and}
  \bibinfo{author}{\bibfnamefont{A.~C.} \bibnamefont{Vincent}},
  \bibinfo{journal}{JCAP} \textbf{\bibinfo{volume}{1302}}, \bibinfo{pages}{011}
  (\bibinfo{year}{2013}), \eprint{1210.2717}.

\bibitem[{\citenamefont{Vernstrom et~al.}(2014)\citenamefont{Vernstrom, Norris,
  Scott, and Wall}}]{Vernstrom:2014uda}
\bibinfo{author}{\bibfnamefont{T.}~\bibnamefont{Vernstrom}},
  \bibinfo{author}{\bibfnamefont{R.~P.} \bibnamefont{Norris}},
  \bibinfo{author}{\bibfnamefont{D.}~\bibnamefont{Scott}}, \bibnamefont{and}
  \bibinfo{author}{\bibfnamefont{J.}~\bibnamefont{Wall}}
  (\bibinfo{year}{2014}), \eprint{1408.4160}.

\bibitem[{\citenamefont{{Ando} and
  {Komatsu}}(2006)}]{ando_komatsu_anisotropy_2006}
\bibinfo{author}{\bibfnamefont{S.}~\bibnamefont{{Ando}}} \bibnamefont{and}
  \bibinfo{author}{\bibfnamefont{E.}~\bibnamefont{{Komatsu}}},
  \bibinfo{journal}{\prd} \textbf{\bibinfo{volume}{73}}, \bibinfo{eid}{023521}
  (\bibinfo{year}{2006}), \eprint{arXiv:astro-ph/0512217}.

\bibitem[{\citenamefont{{Ando} and
  {Komatsu}}(2013)}]{ando_komatsu_anisotropy_2013}
\bibinfo{author}{\bibfnamefont{S.}~\bibnamefont{{Ando}}} \bibnamefont{and}
  \bibinfo{author}{\bibfnamefont{E.}~\bibnamefont{{Komatsu}}},
  \bibinfo{journal}{ArXiv e-prints}  (\bibinfo{year}{2013}),
  \eprint{1301.5901}.

\bibitem[{\citenamefont{{Fornengo} et~al.}(2012)\citenamefont{{Fornengo},
  {Lineros}, {Regis}, and {Taoso}}}]{fornengo_arcade_excess_is_dm2}
\bibinfo{author}{\bibfnamefont{N.}~\bibnamefont{{Fornengo}}},
  \bibinfo{author}{\bibfnamefont{R.}~\bibnamefont{{Lineros}}},
  \bibinfo{author}{\bibfnamefont{M.}~\bibnamefont{{Regis}}}, \bibnamefont{and}
  \bibinfo{author}{\bibfnamefont{M.}~\bibnamefont{{Taoso}}},
  \bibinfo{journal}{JCAP} \textbf{\bibinfo{volume}{3}}, \bibinfo{eid}{033}
  (\bibinfo{year}{2012}), \eprint{1112.4517}.

\bibitem[{\citenamefont{{Stocke} et~al.}(1992)\citenamefont{{Stocke}, {Morris},
  {Weymann}, and {Foltz}}}]{1992ApJ...396..487S}
\bibinfo{author}{\bibfnamefont{J.~T.} \bibnamefont{{Stocke}}},
  \bibinfo{author}{\bibfnamefont{S.~L.} \bibnamefont{{Morris}}},
  \bibinfo{author}{\bibfnamefont{R.~J.} \bibnamefont{{Weymann}}},
  \bibnamefont{and} \bibinfo{author}{\bibfnamefont{C.~B.}
  \bibnamefont{{Foltz}}}, \bibinfo{journal}{\apj}
  \textbf{\bibinfo{volume}{396}}, \bibinfo{pages}{487} (\bibinfo{year}{1992}).

\bibitem[{\citenamefont{{Komatsu} et~al.}(2011)\citenamefont{{Komatsu},
  {Smith}, {Dunkley}, {Bennett}, {Gold}, {Hinshaw}, {Jarosik}, {Larson},
  {Nolta}, {Page} et~al.}}]{2011ApJS..192...18K}
\bibinfo{author}{\bibfnamefont{E.}~\bibnamefont{{Komatsu}}},
  \bibinfo{author}{\bibfnamefont{K.~M.} \bibnamefont{{Smith}}},
  \bibinfo{author}{\bibfnamefont{J.}~\bibnamefont{{Dunkley}}},
  \bibinfo{author}{\bibfnamefont{C.~L.} \bibnamefont{{Bennett}}},
  \bibinfo{author}{\bibfnamefont{B.}~\bibnamefont{{Gold}}},
  \bibinfo{author}{\bibfnamefont{G.}~\bibnamefont{{Hinshaw}}},
  \bibinfo{author}{\bibfnamefont{N.}~\bibnamefont{{Jarosik}}},
  \bibinfo{author}{\bibfnamefont{D.}~\bibnamefont{{Larson}}},
  \bibinfo{author}{\bibfnamefont{M.~R.} \bibnamefont{{Nolta}}},
  \bibinfo{author}{\bibfnamefont{L.}~\bibnamefont{{Page}}},
  \bibnamefont{et~al.}, \bibinfo{journal}{ApJS} \textbf{\bibinfo{volume}{192}},
  \bibinfo{eid}{18} (\bibinfo{year}{2011}), \eprint{1001.4538}.

\bibitem[{\citenamefont{{Cooray} and {Sheth}}(2002)}]{2002PhR...372....1C}
\bibinfo{author}{\bibfnamefont{A.}~\bibnamefont{{Cooray}}} \bibnamefont{and}
  \bibinfo{author}{\bibfnamefont{R.}~\bibnamefont{{Sheth}}},
  \bibinfo{journal}{Physics Reports} \textbf{\bibinfo{volume}{372}},
  \bibinfo{pages}{1} (\bibinfo{year}{2002}), \eprint{astro-ph/0206508}.

\bibitem[{\citenamefont{{Eisenstein} and {Hu}}(1999)}]{1999ApJ...511....5E}
\bibinfo{author}{\bibfnamefont{D.~J.} \bibnamefont{{Eisenstein}}}
  \bibnamefont{and} \bibinfo{author}{\bibfnamefont{W.}~\bibnamefont{{Hu}}},
  \bibinfo{journal}{\apj} \textbf{\bibinfo{volume}{511}}, \bibinfo{pages}{5}
  (\bibinfo{year}{1999}), \eprint{astro-ph/9710252}.

\bibitem[{\citenamefont{{Tinker} et~al.}(2008)\citenamefont{{Tinker},
  {Kravtsov}, {Klypin}, {Abazajian}, {Warren}, {Yepes}, {Gottl{\"o}ber}, and
  {Holz}}}]{2008ApJ...688..709T}
\bibinfo{author}{\bibfnamefont{J.}~\bibnamefont{{Tinker}}},
  \bibinfo{author}{\bibfnamefont{A.~V.} \bibnamefont{{Kravtsov}}},
  \bibinfo{author}{\bibfnamefont{A.}~\bibnamefont{{Klypin}}},
  \bibinfo{author}{\bibfnamefont{K.}~\bibnamefont{{Abazajian}}},
  \bibinfo{author}{\bibfnamefont{M.}~\bibnamefont{{Warren}}},
  \bibinfo{author}{\bibfnamefont{G.}~\bibnamefont{{Yepes}}},
  \bibinfo{author}{\bibfnamefont{S.}~\bibnamefont{{Gottl{\"o}ber}}},
  \bibnamefont{and} \bibinfo{author}{\bibfnamefont{D.~E.}
  \bibnamefont{{Holz}}}, \bibinfo{journal}{\apj}
  \textbf{\bibinfo{volume}{688}}, \bibinfo{pages}{709} (\bibinfo{year}{2008}),
  \eprint{0803.2706}.

\bibitem[{\citenamefont{{Sheth} and
  {Tormen}}(1999)}]{sheth_tormen_mass_function}
\bibinfo{author}{\bibfnamefont{R.~K.} \bibnamefont{{Sheth}}} \bibnamefont{and}
  \bibinfo{author}{\bibfnamefont{G.}~\bibnamefont{{Tormen}}},
  \bibinfo{journal}{MNRAS} \textbf{\bibinfo{volume}{308}}, \bibinfo{pages}{119}
  (\bibinfo{year}{1999}), \eprint{arXiv:astro-ph/9901122}.

\bibitem[{\citenamefont{{Press} and
  {Schechter}}(1974)}]{press_schechter_mass_function}
\bibinfo{author}{\bibfnamefont{W.~H.} \bibnamefont{{Press}}} \bibnamefont{and}
  \bibinfo{author}{\bibfnamefont{P.}~\bibnamefont{{Schechter}}},
  \bibinfo{journal}{ApJ} \textbf{\bibinfo{volume}{187}}, \bibinfo{pages}{425}
  (\bibinfo{year}{1974}).

\bibitem[{\citenamefont{{Navarro} et~al.}(1996)\citenamefont{{Navarro},
  {Frenk}, and {White}}}]{nfw_profile}
\bibinfo{author}{\bibfnamefont{J.~F.} \bibnamefont{{Navarro}}},
  \bibinfo{author}{\bibfnamefont{C.~S.} \bibnamefont{{Frenk}}},
  \bibnamefont{and} \bibinfo{author}{\bibfnamefont{S.~D.~M.}
  \bibnamefont{{White}}}, \bibinfo{journal}{ApJ}
  \textbf{\bibinfo{volume}{462}}, \bibinfo{pages}{563} (\bibinfo{year}{1996}),
  \eprint{arXiv:astro-ph/9508025}.

\bibitem[{\citenamefont{{Gnedin} et~al.}(2011)\citenamefont{{Gnedin},
  {Ceverino}, {Gnedin}, {Klypin}, {Kravtsov}, {Levine}, {Nagai}, and
  {Yepes}}}]{gnedin_adiabatic_contraction}
\bibinfo{author}{\bibfnamefont{O.~Y.} \bibnamefont{{Gnedin}}},
  \bibinfo{author}{\bibfnamefont{D.}~\bibnamefont{{Ceverino}}},
  \bibinfo{author}{\bibfnamefont{N.~Y.} \bibnamefont{{Gnedin}}},
  \bibinfo{author}{\bibfnamefont{A.~A.} \bibnamefont{{Klypin}}},
  \bibinfo{author}{\bibfnamefont{A.~V.} \bibnamefont{{Kravtsov}}},
  \bibinfo{author}{\bibfnamefont{R.}~\bibnamefont{{Levine}}},
  \bibinfo{author}{\bibfnamefont{D.}~\bibnamefont{{Nagai}}}, \bibnamefont{and}
  \bibinfo{author}{\bibfnamefont{G.}~\bibnamefont{{Yepes}}},
  \bibinfo{journal}{ArXiv e-prints}  (\bibinfo{year}{2011}),
  \eprint{1108.5736}.

\bibitem[{\citenamefont{{Klypin} et~al.}(2011)\citenamefont{{Klypin},
  {Trujillo-Gomez}, and {Primack}}}]{klypin_concentration_parameter}
\bibinfo{author}{\bibfnamefont{A.~A.} \bibnamefont{{Klypin}}},
  \bibinfo{author}{\bibfnamefont{S.}~\bibnamefont{{Trujillo-Gomez}}},
  \bibnamefont{and}
  \bibinfo{author}{\bibfnamefont{J.}~\bibnamefont{{Primack}}},
  \bibinfo{journal}{ApJ} \textbf{\bibinfo{volume}{740}}, \bibinfo{eid}{102}
  (\bibinfo{year}{2011}), \eprint{1002.3660}.

\bibitem[{\citenamefont{{Hu} and {Kravtsov}}(2003)}]{2003ApJ...584..702H}
\bibinfo{author}{\bibfnamefont{W.}~\bibnamefont{{Hu}}} \bibnamefont{and}
  \bibinfo{author}{\bibfnamefont{A.~V.} \bibnamefont{{Kravtsov}}},
  \bibinfo{journal}{\apj} \textbf{\bibinfo{volume}{584}}, \bibinfo{pages}{702}
  (\bibinfo{year}{2003}), \eprint{astro-ph/0203169}.

\bibitem[{\citenamefont{{Kamionkowski}
  et~al.}(2010)\citenamefont{{Kamionkowski}, {Koushiappas}, and
  {Kuhlen}}}]{2010PhRvD..81d3532K}
\bibinfo{author}{\bibfnamefont{M.}~\bibnamefont{{Kamionkowski}}},
  \bibinfo{author}{\bibfnamefont{S.~M.} \bibnamefont{{Koushiappas}}},
  \bibnamefont{and} \bibinfo{author}{\bibfnamefont{M.}~\bibnamefont{{Kuhlen}}},
  \bibinfo{journal}{\prd} \textbf{\bibinfo{volume}{81}}, \bibinfo{eid}{043532}
  (\bibinfo{year}{2010}), \eprint{1001.3144}.

\bibitem[{\citenamefont{{Diemand} et~al.}(2008)\citenamefont{{Diemand},
  {Kuhlen}, {Madau}, {Zemp}, {Moore}, {Potter}, and
  {Stadel}}}]{2008Natur.454..735D}
\bibinfo{author}{\bibfnamefont{J.}~\bibnamefont{{Diemand}}},
  \bibinfo{author}{\bibfnamefont{M.}~\bibnamefont{{Kuhlen}}},
  \bibinfo{author}{\bibfnamefont{P.}~\bibnamefont{{Madau}}},
  \bibinfo{author}{\bibfnamefont{M.}~\bibnamefont{{Zemp}}},
  \bibinfo{author}{\bibfnamefont{B.}~\bibnamefont{{Moore}}},
  \bibinfo{author}{\bibfnamefont{D.}~\bibnamefont{{Potter}}}, \bibnamefont{and}
  \bibinfo{author}{\bibfnamefont{J.}~\bibnamefont{{Stadel}}},
  \bibinfo{journal}{\nat} \textbf{\bibinfo{volume}{454}}, \bibinfo{pages}{735}
  (\bibinfo{year}{2008}), \eprint{0805.1244}.

\bibitem[{\citenamefont{{Hooper} et~al.}(2012)\citenamefont{{Hooper},
  {Belikov}, {Jeltema}, {Linden}, {Profumo}, and
  {Slatyer}}}]{hooper_arcade_excess}
\bibinfo{author}{\bibfnamefont{D.}~\bibnamefont{{Hooper}}},
  \bibinfo{author}{\bibfnamefont{A.~V.} \bibnamefont{{Belikov}}},
  \bibinfo{author}{\bibfnamefont{T.~E.} \bibnamefont{{Jeltema}}},
  \bibinfo{author}{\bibfnamefont{T.}~\bibnamefont{{Linden}}},
  \bibinfo{author}{\bibfnamefont{S.}~\bibnamefont{{Profumo}}},
  \bibnamefont{and} \bibinfo{author}{\bibfnamefont{T.~R.}
  \bibnamefont{{Slatyer}}}, \bibinfo{journal}{PRD}
  \textbf{\bibinfo{volume}{86}}, \bibinfo{eid}{103003} (\bibinfo{year}{2012}),
  \eprint{1203.3547}.

\bibitem[{\citenamefont{{Springel} et~al.}(2008)\citenamefont{{Springel},
  {Wang}, {Vogelsberger}, {Ludlow}, {Jenkins}, {Helmi}, {Navarro}, {Frenk}, and
  {White}}}]{2008MNRAS.391.1685S}
\bibinfo{author}{\bibfnamefont{V.}~\bibnamefont{{Springel}}},
  \bibinfo{author}{\bibfnamefont{J.}~\bibnamefont{{Wang}}},
  \bibinfo{author}{\bibfnamefont{M.}~\bibnamefont{{Vogelsberger}}},
  \bibinfo{author}{\bibfnamefont{A.}~\bibnamefont{{Ludlow}}},
  \bibinfo{author}{\bibfnamefont{A.}~\bibnamefont{{Jenkins}}},
  \bibinfo{author}{\bibfnamefont{A.}~\bibnamefont{{Helmi}}},
  \bibinfo{author}{\bibfnamefont{J.~F.} \bibnamefont{{Navarro}}},
  \bibinfo{author}{\bibfnamefont{C.~S.} \bibnamefont{{Frenk}}},
  \bibnamefont{and} \bibinfo{author}{\bibfnamefont{S.~D.~M.}
  \bibnamefont{{White}}}, \bibinfo{journal}{MNRAS}
  \textbf{\bibinfo{volume}{391}}, \bibinfo{pages}{1685} (\bibinfo{year}{2008}),
  \eprint{0809.0898}.

\bibitem[{\citenamefont{{Madau} et~al.}(2008)\citenamefont{{Madau}, {Diemand},
  and {Kuhlen}}}]{2008ApJ...679.1260M}
\bibinfo{author}{\bibfnamefont{P.}~\bibnamefont{{Madau}}},
  \bibinfo{author}{\bibfnamefont{J.}~\bibnamefont{{Diemand}}},
  \bibnamefont{and} \bibinfo{author}{\bibfnamefont{M.}~\bibnamefont{{Kuhlen}}},
  \bibinfo{journal}{\apj} \textbf{\bibinfo{volume}{679}}, \bibinfo{pages}{1260}
  (\bibinfo{year}{2008}), \eprint{0802.2265}.

\bibitem[{\citenamefont{{Diemer} and {Kravtsov}}(2014)}]{2014ApJ...789....1D}
\bibinfo{author}{\bibfnamefont{B.}~\bibnamefont{{Diemer}}} \bibnamefont{and}
  \bibinfo{author}{\bibfnamefont{A.~V.} \bibnamefont{{Kravtsov}}},
  \bibinfo{journal}{\apj} \textbf{\bibinfo{volume}{789}}, \bibinfo{eid}{1}
  (\bibinfo{year}{2014}), \eprint{1401.1216}.

\bibitem[{\citenamefont{{Gao} et~al.}(2012{\natexlab{a}})\citenamefont{{Gao},
  {Frenk}, {Jenkins}, {Springel}, and {White}}}]{2012MNRAS.419.1721G}
\bibinfo{author}{\bibfnamefont{L.}~\bibnamefont{{Gao}}},
  \bibinfo{author}{\bibfnamefont{C.~S.} \bibnamefont{{Frenk}}},
  \bibinfo{author}{\bibfnamefont{A.}~\bibnamefont{{Jenkins}}},
  \bibinfo{author}{\bibfnamefont{V.}~\bibnamefont{{Springel}}},
  \bibnamefont{and} \bibinfo{author}{\bibfnamefont{S.~D.~M.}
  \bibnamefont{{White}}}, \bibinfo{journal}{MNRAS}
  \textbf{\bibinfo{volume}{419}}, \bibinfo{pages}{1721}
  (\bibinfo{year}{2012}{\natexlab{a}}), \eprint{1107.1916}.

\bibitem[{\citenamefont{{Gao} et~al.}(2012{\natexlab{b}})\citenamefont{{Gao},
  {Navarro}, {Frenk}, {Jenkins}, {Springel}, and
  {White}}}]{2012MNRAS.425.2169G}
\bibinfo{author}{\bibfnamefont{L.}~\bibnamefont{{Gao}}},
  \bibinfo{author}{\bibfnamefont{J.~F.} \bibnamefont{{Navarro}}},
  \bibinfo{author}{\bibfnamefont{C.~S.} \bibnamefont{{Frenk}}},
  \bibinfo{author}{\bibfnamefont{A.}~\bibnamefont{{Jenkins}}},
  \bibinfo{author}{\bibfnamefont{V.}~\bibnamefont{{Springel}}},
  \bibnamefont{and} \bibinfo{author}{\bibfnamefont{S.~D.~M.}
  \bibnamefont{{White}}}, \bibinfo{journal}{MNRAS}
  \textbf{\bibinfo{volume}{425}}, \bibinfo{pages}{2169}
  (\bibinfo{year}{2012}{\natexlab{b}}), \eprint{1201.1940}.

\bibitem[{\citenamefont{{S{\'a}nchez-Conde} and
  {Prada}}(2014)}]{2014MNRAS.442.2271S}
\bibinfo{author}{\bibfnamefont{M.~A.} \bibnamefont{{S{\'a}nchez-Conde}}}
  \bibnamefont{and} \bibinfo{author}{\bibfnamefont{F.}~\bibnamefont{{Prada}}},
  \bibinfo{journal}{MNRAS} \textbf{\bibinfo{volume}{442}},
  \bibinfo{pages}{2271} (\bibinfo{year}{2014}), \eprint{1312.1729}.

\bibitem[{\citenamefont{{Crocker} et~al.}(2010)\citenamefont{{Crocker},
  {Jones}, {Melia}, {Ott}, and {Protheroe}}}]{crocker_minimum_50muG}
\bibinfo{author}{\bibfnamefont{R.~M.} \bibnamefont{{Crocker}}},
  \bibinfo{author}{\bibfnamefont{D.~I.} \bibnamefont{{Jones}}},
  \bibinfo{author}{\bibfnamefont{F.}~\bibnamefont{{Melia}}},
  \bibinfo{author}{\bibfnamefont{J.}~\bibnamefont{{Ott}}}, \bibnamefont{and}
  \bibinfo{author}{\bibfnamefont{R.~J.} \bibnamefont{{Protheroe}}},
  \bibinfo{journal}{Nat} \textbf{\bibinfo{volume}{463}}, \bibinfo{pages}{65}
  (\bibinfo{year}{2010}), \eprint{1001.1275}.

\bibitem[{\citenamefont{{Carilli} and {Taylor}}(2002)}]{2002ARA&A..40..319C}
\bibinfo{author}{\bibfnamefont{C.~L.} \bibnamefont{{Carilli}}}
  \bibnamefont{and} \bibinfo{author}{\bibfnamefont{G.~B.}
  \bibnamefont{{Taylor}}}, \bibinfo{journal}{Annual Review of Astronomy and
  Astrophysics} \textbf{\bibinfo{volume}{40}}, \bibinfo{pages}{319}
  (\bibinfo{year}{2002}), \eprint{astro-ph/0110655}.

\bibitem[{\citenamefont{{Bonafede} et~al.}(2009)\citenamefont{{Bonafede},
  {Feretti}, {Giovannini}, {Govoni}, {Murgia}, {Taylor}, {Ebeling}, {Allen},
  {Gentile}, and {Pihlstr{\"o}m}}}]{2009A&A...503..707B}
\bibinfo{author}{\bibfnamefont{A.}~\bibnamefont{{Bonafede}}},
  \bibinfo{author}{\bibfnamefont{L.}~\bibnamefont{{Feretti}}},
  \bibinfo{author}{\bibfnamefont{G.}~\bibnamefont{{Giovannini}}},
  \bibinfo{author}{\bibfnamefont{F.}~\bibnamefont{{Govoni}}},
  \bibinfo{author}{\bibfnamefont{M.}~\bibnamefont{{Murgia}}},
  \bibinfo{author}{\bibfnamefont{G.~B.} \bibnamefont{{Taylor}}},
  \bibinfo{author}{\bibfnamefont{H.}~\bibnamefont{{Ebeling}}},
  \bibinfo{author}{\bibfnamefont{S.}~\bibnamefont{{Allen}}},
  \bibinfo{author}{\bibfnamefont{G.}~\bibnamefont{{Gentile}}},
  \bibnamefont{and}
  \bibinfo{author}{\bibfnamefont{Y.}~\bibnamefont{{Pihlstr{\"o}m}}},
  \bibinfo{journal}{AAP} \textbf{\bibinfo{volume}{503}}, \bibinfo{pages}{707}
  (\bibinfo{year}{2009}), \eprint{0905.3552}.

\bibitem[{\citenamefont{{Br{\"u}ggen} et~al.}(2012)\citenamefont{{Br{\"u}ggen},
  {Bykov}, {Ryu}, and {R{\"o}ttgering}}}]{2012SSRv..166..187B}
\bibinfo{author}{\bibfnamefont{M.}~\bibnamefont{{Br{\"u}ggen}}},
  \bibinfo{author}{\bibfnamefont{A.}~\bibnamefont{{Bykov}}},
  \bibinfo{author}{\bibfnamefont{D.}~\bibnamefont{{Ryu}}}, \bibnamefont{and}
  \bibinfo{author}{\bibfnamefont{H.}~\bibnamefont{{R{\"o}ttgering}}},
  \bibinfo{journal}{Space Science Reviews} \textbf{\bibinfo{volume}{166}},
  \bibinfo{pages}{187} (\bibinfo{year}{2012}), \eprint{1107.5223}.

\bibitem[{\citenamefont{{Clarke} and {Ensslin}}(2006)}]{2006AJ....131.2900C}
\bibinfo{author}{\bibfnamefont{T.~E.} \bibnamefont{{Clarke}}} \bibnamefont{and}
  \bibinfo{author}{\bibfnamefont{T.~A.} \bibnamefont{{Ensslin}}},
  \bibinfo{journal}{AJ} \textbf{\bibinfo{volume}{131}}, \bibinfo{pages}{2900}
  (\bibinfo{year}{2006}), \eprint{astro-ph/0603166}.

\bibitem[{\citenamefont{{van Weeren} et~al.}(2010)\citenamefont{{van Weeren},
  {R{\"o}ttgering}, {Br{\"u}ggen}, and {Hoeft}}}]{2010Sci...330..347V}
\bibinfo{author}{\bibfnamefont{R.~J.} \bibnamefont{{van Weeren}}},
  \bibinfo{author}{\bibfnamefont{H.~J.~A.} \bibnamefont{{R{\"o}ttgering}}},
  \bibinfo{author}{\bibfnamefont{M.}~\bibnamefont{{Br{\"u}ggen}}},
  \bibnamefont{and} \bibinfo{author}{\bibfnamefont{M.}~\bibnamefont{{Hoeft}}},
  \bibinfo{journal}{Science} \textbf{\bibinfo{volume}{330}},
  \bibinfo{pages}{347} (\bibinfo{year}{2010}), \eprint{1010.4306}.

\bibitem[{\citenamefont{{Beck} and {Krause}}(2005)}]{2005AN....326..414B}
\bibinfo{author}{\bibfnamefont{R.}~\bibnamefont{{Beck}}} \bibnamefont{and}
  \bibinfo{author}{\bibfnamefont{M.}~\bibnamefont{{Krause}}},
  \bibinfo{journal}{Astronomische Nachrichten} \textbf{\bibinfo{volume}{326}},
  \bibinfo{pages}{414} (\bibinfo{year}{2005}), \eprint{astro-ph/0507367}.

\bibitem[{\citenamefont{{Blasi} and {Amato}}(2012)}]{2012JCAP...01..010B}
\bibinfo{author}{\bibfnamefont{P.}~\bibnamefont{{Blasi}}} \bibnamefont{and}
  \bibinfo{author}{\bibfnamefont{E.}~\bibnamefont{{Amato}}},
  \bibinfo{journal}{JCAP} \textbf{\bibinfo{volume}{1}}, \bibinfo{eid}{010}
  (\bibinfo{year}{2012}), \eprint{1105.4521}.

\bibitem[{\citenamefont{{Fouka} and {Ouichaoui}}(2013)}]{2013RAA....13..680F}
\bibinfo{author}{\bibfnamefont{M.}~\bibnamefont{{Fouka}}} \bibnamefont{and}
  \bibinfo{author}{\bibfnamefont{S.}~\bibnamefont{{Ouichaoui}}},
  \bibinfo{journal}{Research in Astronomy and Astrophysics}
  \textbf{\bibinfo{volume}{13}}, \bibinfo{eid}{680} (\bibinfo{year}{2013}),
  \eprint{1301.6908}.

\bibitem[{\citenamefont{Geringer-Sameth and
  Koushiappas}(2011)}]{GeringerSameth:2011iw}
\bibinfo{author}{\bibfnamefont{A.}~\bibnamefont{Geringer-Sameth}}
  \bibnamefont{and} \bibinfo{author}{\bibfnamefont{S.~M.}
  \bibnamefont{Koushiappas}}, \bibinfo{journal}{Phys.Rev.Lett.}
  \textbf{\bibinfo{volume}{107}}, \bibinfo{pages}{241303}
  (\bibinfo{year}{2011}), \eprint{1108.2914}.

\bibitem[{\citenamefont{Ackermann et~al.}(2014)}]{Ackermann:2013yva}
\bibinfo{author}{\bibfnamefont{M.}~\bibnamefont{Ackermann}}
  \bibnamefont{et~al.} (\bibinfo{collaboration}{Fermi-LAT Collaboration}),
  \bibinfo{journal}{Phys.Rev.} \textbf{\bibinfo{volume}{D89}},
  \bibinfo{pages}{042001} (\bibinfo{year}{2014}), \eprint{1310.0828}.

\bibitem[{\citenamefont{Geringer-Sameth
  et~al.}(2014)\citenamefont{Geringer-Sameth, Koushiappas, and
  Walker}}]{Geringer-Sameth:2014qqa}
\bibinfo{author}{\bibfnamefont{A.}~\bibnamefont{Geringer-Sameth}},
  \bibinfo{author}{\bibfnamefont{S.~M.} \bibnamefont{Koushiappas}},
  \bibnamefont{and} \bibinfo{author}{\bibfnamefont{M.~G.} \bibnamefont{Walker}}
  (\bibinfo{year}{2014}), \eprint{1410.2242}.

\bibitem[{\citenamefont{Anderson}(2014)}]{Anderson_Symposium}
\bibinfo{author}{\bibfnamefont{B.}~\bibnamefont{Anderson}}
  (\bibinfo{year}{2014}), \eprint{5th Fermi-LAT Symposium}.

\bibitem[{\citenamefont{{Linden} et~al.}(2011)\citenamefont{{Linden}, {Hooper},
  and {Yusef-Zadeh}}}]{linden_filamentary_arcs}
\bibinfo{author}{\bibfnamefont{T.}~\bibnamefont{{Linden}}},
  \bibinfo{author}{\bibfnamefont{D.}~\bibnamefont{{Hooper}}}, \bibnamefont{and}
  \bibinfo{author}{\bibfnamefont{F.}~\bibnamefont{{Yusef-Zadeh}}},
  \bibinfo{journal}{ApJ} \textbf{\bibinfo{volume}{741}}, \bibinfo{eid}{95}
  (\bibinfo{year}{2011}), \eprint{1106.5493}.

\bibitem[{\citenamefont{Lacroix et~al.}(2014)\citenamefont{Lacroix, Boehm, and
  Silk}}]{Lacroix:2014eea}
\bibinfo{author}{\bibfnamefont{T.}~\bibnamefont{Lacroix}},
  \bibinfo{author}{\bibfnamefont{C.}~\bibnamefont{Boehm}}, \bibnamefont{and}
  \bibinfo{author}{\bibfnamefont{J.}~\bibnamefont{Silk}},
  \bibinfo{journal}{Phys.Rev.} \textbf{\bibinfo{volume}{D90}},
  \bibinfo{pages}{043508} (\bibinfo{year}{2014}), \eprint{1403.1987}.

\bibitem[{\citenamefont{Bergstrom et~al.}(2013)\citenamefont{Bergstrom,
  Bringmann, Cholis, Hooper, and Weniger}}]{Bergstrom:2013jra}
\bibinfo{author}{\bibfnamefont{L.}~\bibnamefont{Bergstrom}},
  \bibinfo{author}{\bibfnamefont{T.}~\bibnamefont{Bringmann}},
  \bibinfo{author}{\bibfnamefont{I.}~\bibnamefont{Cholis}},
  \bibinfo{author}{\bibfnamefont{D.}~\bibnamefont{Hooper}}, \bibnamefont{and}
  \bibinfo{author}{\bibfnamefont{C.}~\bibnamefont{Weniger}},
  \bibinfo{journal}{Phys.Rev.Lett.} \textbf{\bibinfo{volume}{111}},
  \bibinfo{pages}{171101} (\bibinfo{year}{2013}), \eprint{1306.3983}.

\bibitem[{\citenamefont{Fairbairn and Grothaus}(2014)}]{Fairbairn:2014nga}
\bibinfo{author}{\bibfnamefont{M.}~\bibnamefont{Fairbairn}} \bibnamefont{and}
  \bibinfo{author}{\bibfnamefont{P.}~\bibnamefont{Grothaus}}
  (\bibinfo{year}{2014}), \eprint{1407.4849}.

\bibitem[{\citenamefont{Slatyer et~al.}(2009)\citenamefont{Slatyer,
  Padmanabhan, and Finkbeiner}}]{Slatyer:2009yq}
\bibinfo{author}{\bibfnamefont{T.~R.} \bibnamefont{Slatyer}},
  \bibinfo{author}{\bibfnamefont{N.}~\bibnamefont{Padmanabhan}},
  \bibnamefont{and} \bibinfo{author}{\bibfnamefont{D.~P.}
  \bibnamefont{Finkbeiner}}, \bibinfo{journal}{Phys.Rev.}
  \textbf{\bibinfo{volume}{D80}}, \bibinfo{pages}{043526}
  (\bibinfo{year}{2009}), \eprint{0906.1197}.

\bibitem[{\citenamefont{Ferrari et~al.}(2008)\citenamefont{Ferrari, Govoni,
  Schindler, Bykov, and Rephaeli}}]{Ferrari:2008jr}
\bibinfo{author}{\bibfnamefont{C.}~\bibnamefont{Ferrari}},
  \bibinfo{author}{\bibfnamefont{F.}~\bibnamefont{Govoni}},
  \bibinfo{author}{\bibfnamefont{S.}~\bibnamefont{Schindler}},
  \bibinfo{author}{\bibfnamefont{A.}~\bibnamefont{Bykov}}, \bibnamefont{and}
  \bibinfo{author}{\bibfnamefont{Y.}~\bibnamefont{Rephaeli}},
  \bibinfo{journal}{Space Sci.Rev.} \textbf{\bibinfo{volume}{134}},
  \bibinfo{pages}{93} (\bibinfo{year}{2008}), \eprint{0801.0985}.

\bibitem[{\citenamefont{{Brunetti} et~al.}(2008)\citenamefont{{Brunetti},
  {Giacintucci}, {Cassano}, {Lane}, {Dallacasa}, {Venturi}, {Kassim}, {Setti},
  {Cotton}, and {Markevitch}}}]{2008Natur.455..944B}
\bibinfo{author}{\bibfnamefont{G.}~\bibnamefont{{Brunetti}}},
  \bibinfo{author}{\bibfnamefont{S.}~\bibnamefont{{Giacintucci}}},
  \bibinfo{author}{\bibfnamefont{R.}~\bibnamefont{{Cassano}}},
  \bibinfo{author}{\bibfnamefont{W.}~\bibnamefont{{Lane}}},
  \bibinfo{author}{\bibfnamefont{D.}~\bibnamefont{{Dallacasa}}},
  \bibinfo{author}{\bibfnamefont{T.}~\bibnamefont{{Venturi}}},
  \bibinfo{author}{\bibfnamefont{N.~E.} \bibnamefont{{Kassim}}},
  \bibinfo{author}{\bibfnamefont{G.}~\bibnamefont{{Setti}}},
  \bibinfo{author}{\bibfnamefont{W.~D.} \bibnamefont{{Cotton}}},
  \bibnamefont{and}
  \bibinfo{author}{\bibfnamefont{M.}~\bibnamefont{{Markevitch}}},
  \bibinfo{journal}{\nat} \textbf{\bibinfo{volume}{455}}, \bibinfo{pages}{944}
  (\bibinfo{year}{2008}), \eprint{0810.4288}.

\bibitem[{\citenamefont{{Brunetti} et~al.}(2004)\citenamefont{{Brunetti},
  {Blasi}, {Cassano}, and {Gabici}}}]{2004MNRAS.350.1174B}
\bibinfo{author}{\bibfnamefont{G.}~\bibnamefont{{Brunetti}}},
  \bibinfo{author}{\bibfnamefont{P.}~\bibnamefont{{Blasi}}},
  \bibinfo{author}{\bibfnamefont{R.}~\bibnamefont{{Cassano}}},
  \bibnamefont{and} \bibinfo{author}{\bibfnamefont{S.}~\bibnamefont{{Gabici}}},
  \bibinfo{journal}{MNRAS} \textbf{\bibinfo{volume}{350}},
  \bibinfo{pages}{1174} (\bibinfo{year}{2004}), \eprint{astro-ph/0312482}.

\bibitem[{\citenamefont{{Brunetti} and {Blasi}}(2005)}]{2005MNRAS.363.1173B}
\bibinfo{author}{\bibfnamefont{G.}~\bibnamefont{{Brunetti}}} \bibnamefont{and}
  \bibinfo{author}{\bibfnamefont{P.}~\bibnamefont{{Blasi}}},
  \bibinfo{journal}{MNRAS} \textbf{\bibinfo{volume}{363}},
  \bibinfo{pages}{1173} (\bibinfo{year}{2005}), \eprint{astro-ph/0508100}.

\bibitem[{\citenamefont{Fang and Linden}(2015)}]{fang_linden_15}
\bibinfo{author}{\bibfnamefont{K.}~\bibnamefont{Fang}} \bibnamefont{and}
  \bibinfo{author}{\bibfnamefont{T.}~\bibnamefont{Linden}}
  (\bibinfo{year}{2015}), \eprint{To Be Submitted.}

\bibitem[{\citenamefont{{Cavaliere} and
  {Fusco-Femiano}}(1976)}]{1976A&A....49..137C}
\bibinfo{author}{\bibfnamefont{A.}~\bibnamefont{{Cavaliere}}} \bibnamefont{and}
  \bibinfo{author}{\bibfnamefont{R.}~\bibnamefont{{Fusco-Femiano}}},
  \bibinfo{journal}{A\&A} \textbf{\bibinfo{volume}{49}}, \bibinfo{pages}{137}
  (\bibinfo{year}{1976}).

\end{thebibliography}

\end{document}